%% file: ANA-HMBS-2024-66-PAPER.tex
\pdfoutput=1
\documentclass[PAPER, atlasdraft=false, cernpreprint, UKenglish, texmf, orcidlogo]{atlasdoc}
\usepackage[subfigure]{atlaspackage}%
\usepackage{multirow}
\usepackage{atlaspackage}
\usepackage{atlasbiblatex}

\usepackage{atlasphysics}
\graphicspath{{logos/}{figures/}}

\usepackage{atlasbsm}
\usepackage{ANA-HMBS-2024-66-PAPER-defs}

\AtlasTitle{Search for long-lived charginos and $\tau$-sleptons using final states with a disappearing track in $pp$ collisions at $\sqrt{s} = 13$ TeV with the ATLAS detector}

\AtlasAbstract{%
This paper reports a search for decays of long-lived charginos or $\tau$-sleptons to final states containing a short disappearing track, a single high-energy jet, and missing transverse momentum. The search uses 137\,fb$^{-1}$ of data from 13 TeV proton--proton collisions recorded by the ATLAS detector during Run~2 of the LHC. Multiple search regions are defined, all requiring the presence of a track reconstructed from either three or four measurements in the innermost layers of the ATLAS detector. Regions with tracks having only three measurements are further characterised by the absence or presence of a low-energy charged pion reconstructed using a dedicated algorithm, leveraging machine learning. Data-driven methods are used to estimate the background contributions in the search regions. No significant excesses are found and 95\% CL lower limits are placed on the masses of charginos and $\tau$-sleptons in the lifetime range $0.01{-}10$ ns. Observed (expected) mass limits of up to 225 GeV (250 GeV) are set for pure-higgsino charginos in scenarios with lifetimes below 0.03 ns, where the electroweakino mass splitting is entirely due to loop corrections involving the Standard Model bosons, and up to 720 GeV (840 GeV) for charginos with a lifetime of around 1 ns. For wino production, charginos with masses up to 880 GeV (1020 GeV) are excluded for lifetimes of around 1 ns. For $\tau$-sleptons with lifetimes of around 1 ns, masses are excluded up to 320 GeV (390 GeV) in Constrained Minimal Supersymmetric Standard Model scenarios and 300 GeV (380 GeV) in Gauge-Mediated Supersymmetry-Breaking scenarios.}

\AtlasRefCode{HMBS-2024-66}

\PreprintIdNumber{CERN-EP-2026-044}

\AtlasJournal{JHEP}
\AtlasJournalRef{JHEP 07 (2026) 152}
\AtlasDOI{10.1007/JHEP07(2026)152}

\AtlasCoverEgroupAnalysisTeam{}

\hypersetup{pdftitle={ATLAS document},pdfauthor={The ATLAS Collaboration}}

\begin{document}

\maketitle

\tableofcontents

\section{Introduction}
%
%

% The next lines are included from the .//sections/intro.tex input file
Supersymmetry~(SUSY)~\cite{Golfand:1971iw,Volkov:1973ix,Wess:1974tw,Wess:1974jb,Ferrara:1974pu,Salam:1974ig} is a theoretical framework based on a space-time symmetry that predicts the existence of
a partner for each Standard Model~(SM) particle. These partner particles have the same quantum numbers as their SM counterparts, but differ by one half unit of spin. SUSY offers a solution to the hierarchy problem~\cite{Sakai:1981gr,Dimopoulos:1981yj,Ibanez:1981yh,Dimopoulos:1981zb} if some of the partner particles have masses around the electroweak scale.
The supersymmetric partners are named to reflect their SM counterparts: for example, the $\tau$-slepton is the SUSY partner of the $\tau$-lepton.
The supersymmetric partners of the electroweak gauge bosons and the Higgs bosons, collectively referred to as electroweakinos, consist of the bino and winos and the higgsinos, respectively.
These states mix to form the mass eigenstates:
the bino, neutral wino, and neutral higgsinos mix to form the neutralinos ($\tilde{\chi}_{1,2,3,4}^{0}$), while the charged wino and charged higgsinos mix to form charginos ($\tilde{\chi}_{1,2}^{\pm}$), where the subscripts denote increasing mass.
The superpartners of the third-generation charged leptons ($\tau_{\mathrm{L}}, \tau_{\mathrm{R}}$) are referred to as $\tau$-sleptons ($\stau_{\mathrm{L}}, \stau_{\mathrm{R}}$) and mix to form two mass eigenstates $\stau_{1}$ and $\stau_{2}$, where $\stau_{1}$ is the lighter of the two and is henceforth referred to as the $\tau$-slepton ($\stau$). $R$-parity-conserving supersymmetry~\cite{Farrar:1978xj} addresses several limitations of the SM, providing a natural candidate for dark matter~(DM) in the form of the lightest supersymmetric particle~(LSP)~\cite{Goldberg:1983nd,Ellis:1983ew}. In SUSY scenarios with very heavy, strongly interacting superpartners, the direct production cross-section of electroweakinos and sleptons in high-energy proton-proton (\pp) collisions at the Large Hadron Collider~(LHC) can significantly exceed that of strongly interacting superpartners.

In scenarios where the \ninoone and \chinoonepm are wino-like and all other SUSY particles are decoupled, the mass difference, $\mdiffchi$, is small ($\sim$100\,\MeV) and arises from SM radiative corrections~\cite{Ibe:2012sx,Fukuda:2017jmk}.
Anomaly-mediated supersymmetry-breaking~(AMSB) models~\cite{Giudice:1998xp,Randall:1998uk} naturally predict such mass splittings and give rise to a wino-like LSP.
In these scenarios, the mass-splitting between the charged wino and neutral wino is suppressed at tree level by the approximate custodial symmetry,
and has been calculated at the two-loop level to be approximately $160\,\MeV$~\cite{Ibe:2012sx}, corresponding to a chargino lifetime
of $\tau(\chinoonepm) \sim 0.2$~ns and a decay length of $c\tau(\chinoonepm) \sim 60$~mm.
Furthermore, a number of \enquote{natural} SUSY models~\cite{Papucci:2011wy,Barbieri:1987fn,deCarlos:1993yy}, where the \ninotwo is no longer decoupled, predict a light higgsino LSP with a mass near the electroweak scale.
In these scenarios, the higgsino mass parameter $|\mu|$ is small compared to the other electroweakino mass parameters, which could mean
higgsinos are discoverable at the LHC. While the charged and neutral higgsino states are mass degenerate at tree level, higher-order SM loop corrections generate a mass splitting of approximately $300\,\MeV$.
For chargino masses between $91\,\GeV$ and $1000\,\GeV$, the mass splitting ranges approximately from $280\,\MeV$ to $350\,\MeV$~\cite{Thomas:1998wy},
corresponding to chargino lifetimes of $\tau(\chinoonepm) \sim 0.026{-}0.048$~ns ($c\tau(\chinoonepm) \sim 7{-}14$~mm)~\cite{Fukuda:2017jmk}. Due to the small mass splitting, the \chinoonepm is assumed to decay via $\chinoonepm \rightarrow \pi^\pm\ninoone$ with a branching fraction of 100\% in the wino scenarios and 95\% in the higgsino scenarios~\cite{Nagata:2014wma}. The neutralino masses are set such that $m(\ninotwo) = m(\ninoone)$, hence the \ninotwo cannot decay to the \chinoonepm or \ninoone and is treated as a stable particle that escapes the detector, similarly to the \ninoone.

Models featuring light $\tau$-sleptons can yield a dark-matter relic density consistent with cosmological observations~\cite{Vasquez:2011} and, more generally, light sleptons can participate in the co-annihilation of neutralinos~\cite{Belanger:2004ag,King:2007vh}. Long-lived $\tau$-sleptons arise naturally in co-annihilation dark-matter models~\cite{Ellis:1998kh,Jittoh:2005pq,Kaneko:2008re} and in gauge-mediated supersymmetry-breaking~(GMSB)
models~\cite{Dine:1981gu,AlvarezGaume:1981wy,Nappi:1982hm}.
Two simplified SUSY models are considered for $\tau$-sleptons in this paper. The first is inspired by the Constrained Minimal Supersymmetric Standard Model (CMSSM)~\cite{Chamseddine:1982jx,Barbieri:1982eh,Kane:1993td}, where a long-lived $\tau$-slepton can arise from a small (${\sim}2$\,GeV) mass-splitting between the $\tau$-slepton and the $\ninoone$ LSP, CP-violating phases, and the mixing angle between the left- and right-handed $\tau$-slepton states. The CMSSM-inspired scenarios are parameterised by the $\tau$-slepton mass and lifetime, and the $\tau$-slepton decays with 100\% branching fraction as $\stau\rightarrow\tau\ninoone$.
The second scenario is inspired by GMSB models, where a long-lived $\tau$-slepton can arise from small couplings to its decay products. The LSP is the almost massless gravitino ($\tilde{G}$), the superpartner of a proposed graviton. This model is parameterised by the $\tau$-slepton mass and lifetime, and assumes a 100\% branching fraction for $\stau\rightarrow\tau\gravino$.

In all considered scenarios, the charged SUSY particle (wino, higgsino or $\tau$-slepton) can be produced with large momentum and live long enough to traverse multiple layers of the ATLAS pixel detector before decaying. In models containing charginos, the chargino deposits energy in the innermost tracking layers before it decays and can be reconstructed as a short track if at least three layers of the tracking detector are hit.
The weakly interacting LSP will escape detection and lead to missing transverse momentum ($\vec{p}_\text{T}^\text{miss}$) and the charged pion from the chargino decay has a very low momentum and is not reconstructed as a track by existing algorithms, this gives rise to the characteristic \enquote{disappearing track} signature where the track from the chargino stops a short distance into the detector. The algorithms used to reconstruct the short tracks, referred to as tracklets, need energy deposits (hits) in at least three pixel layers. Therefore successful reconstruction requires charginos to have transverse path lengths that at least exceed the radius of the third pixel layer at  $\sim$88.5~mm.
This makes the higgsino-like scenarios considerably more challenging than the wino-like models from an experimental perspective because of the extremely short chargino lifetime predicted in the higgsino models. A similar scenario occurs in the case of $\tau$-slepton production, where a long-lived $\tau$-slepton particle will traverse the inner layers of the pixel detector before decaying into a $\tau$-lepton and the LSP. The LSP escapes the detector unobserved, resulting in missing transverse momentum, and the $\tau$-leptons evade standard reconstruction techniques, again resulting in a signature containing a disappearing track and missing transverse momentum.

In addition to the SUSY-specific wino and higgsino models studied in this paper, the disappearing-track signature is typical of a large class of DM models that predict a DM thermal relic from a massive particle with only electroweak gauge interactions~\cite{Cirelli:2005uq}.
The wino and higgsino models probed in this search are part of a more generic class of models containing spin-1/2 particles transforming under SU(2) symmetry, which give rise to nearly mass-degenerate doublets ($\chinoonepm$, $\ninoone$) and triplets ($\chinoonepm$, $\ninotwo$, $\ninoone$) of new particles, respectively. In both cases, the $\ninoone$ is a DM candidate, while the $\chinoonepm$ gives rise to the disappearing-track signature. These DM models have gained interest in the wider community as an important area of study at future colliders~\cite{CidVidal:2018eel,Saito:2019rtg,Capdevilla:2021fmj,Capdevilla:2024bwt}.

\begin{figure}[t!]
\begin{center}
\subfigure[]{
\includegraphics[height=0.25\textwidth]{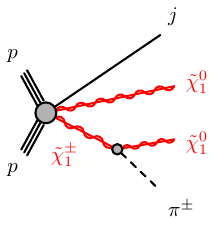}
\label{fig:intro:feynman:c1n1}
}
\subfigure[]{
\includegraphics[height=0.25\textwidth]{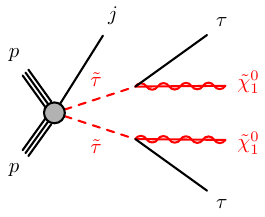}
\label{fig:intro:feynman:ststCMSSM}
}
\subfigure[]{
\includegraphics[height=0.25\textwidth]{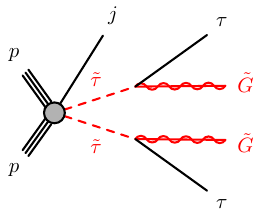}
\label{fig:intro:feynman:ststGMSB}
}
\end{center}
\vspace{-0.75cm}
\caption{Representative signal diagrams for \protect\subref{fig:intro:feynman:c1n1} the electroweak production of $\chinoonepm\ninoone$,
and $\tau$-slepton pair production in \protect\subref{fig:intro:feynman:ststCMSSM} CMSSM-inspired or \protect\subref{fig:intro:feynman:ststGMSB} GMSB-inspired SUSY models.
The signal signature consists of a long-lived \chinoonepm or $\tau$-slepton, missing transverse momentum, and quarks or gluons, which are observed as jets ($j$). When considering the electroweakino scenarios, the production of $\chinoonep\chinoonem$ is also considered, as is $\chinoonepm\ninotwo$ specifically in the higgsino scenario.}
\label{fig:SUSYFeynman}
\end{figure}

This paper targets two electroweak production processes, the production of charginos and neutralinos, and the production of $\tau$-slepton pairs in CMSSM- and GMSB-inspired scenarios, as shown in Figure~\ref{fig:SUSYFeynman}, in the context of simplified SUSY models~\cite{Alwall:2008ve,Alwall:2008ag,Alves:2011wf}.
In all scenarios, the charged SUSY partner is long-lived and can be reconstructed from hits in the ATLAS pixel detector.
A high-momentum jet from initial-state radiation~(ISR) is required in order to ensure that significant missing transverse momentum is available for event triggering.

Previous ATLAS searches for long-lived supersymmetric charged particles searched for events with a disappearing track with at least four pixel hits~\cite{SUSY-2018-19,SUSY-2016-06,ATL-PHYS-PUB-2017-019}. Their results benefited from the addition of a new innermost pixel tracking layer installed during the LHC long shutdown between Run~1 and Run~2. The extra layer of pixels enabled the reconstruction of tracks shorter than those in the previous Run-1 analysis~\cite{SUSY-2013-01} and enhanced the sensitivity to shorter charged-particle lifetimes. The Run-2 ATLAS search excluded wino-like charginos with lifetimes of $0.2\,$ns for masses up to $460\,\GeV$, and higgsino-like charginos with masses up to $152\,\GeV$~\cite{SUSY-2018-19}. Searches performed by the CMS Collaboration excluded wino-like charginos with masses below $474\,\GeV$ and a lifetime of $0.2\,$ns~\cite{CMS-EXO-19-010}.
Both the ATLAS and CMS collaborations have performed searches for stable charginos and $\tau$-sleptons using ionisation-loss or time-of-flight techniques, excluding charginos with masses smaller than $1.3\,\TeV$ for lifetimes exceeding $100\,$ns, and $\tau$-sleptons with masses $200{-}560\,\GeV$ for lifetimes of $10\,$ns~\cite{HMBS-2024-68,CMS-EXO-18-002}.

In order to improve on these results, significant developments were made in the tracking techniques necessary for shorter lifetime scenarios.
This search employs a dedicated tracking algorithm to reconstruct tracklets with as few as three pixel hits.
Furthermore, to enhance sensitivity to the electroweakino production scenario, the charged pion from the decay of the $\chinoonepm$ is targeted with a specially optimised reconstruction algorithm.
Due to the small mass-splitting between the $\chinoonepm$ and $\ninoone$, the charged pion has extremely low momentum, ranging from $250$ to $350\,\MeV$ for the \chinoonepm lifetimes considered, and typically has large impact parameters measured relative to the interaction region.
In this paper, the sensitivity to charginos with wino and higgsino lifetimes motivated by SM loop corrections is significantly improved due
to new analysis methods including updated signal region selection criteria, the previously mentioned reconstruction algorithms, and improved track-quality requirements.
The new track-quality criteria requires the disappearing track not to be associated with any energy deposits in the calorimeter.
This significantly enhances the rejection of dominant backgrounds
and accounts for two-thirds of the total improvement in the mass reach of the search in the electroweak wino-like production channel.

The paper is structured as follows: Section~\ref{sec:detector} introduces the ATLAS detector apparatus; Section~\ref{sec:DataAndMC} describes the dataset and the Monte Carlo simulation samples used; Section~\ref{sec:Reco} discusses the object-level event reconstruction, while Section~\ref{sec:Analysis} describes the signal selection optimisation and Section~\ref{sec:bkgs} describes the background estimation techniques; Section~\ref{sec:Systs} provides an overview of the systematic uncertainties considered, with Section~\ref{sec:result} presenting the results and interpretation thereof; and finally, Section~\ref{sec:conclusion} summarises the conclusions.

% End of text imported from the .//sections/intro.tex input file

%
\section{ATLAS detector}
\label{sec:detector}
%

% The next lines are included from the .//sections/detector.tex input file
%
\newcommand{\AtlasCoordFootnote}{%
ATLAS uses a right-handed coordinate system with its origin at the nominal interaction point (IP)
in the centre of the detector and the \(z\)-axis along the beam pipe.
The \(x\)-axis points from the IP to the centre of the LHC ring,
and the \(y\)-axis points upwards.
Polar coordinates \((r,\phi)\) are used in the transverse plane,
\(\phi\) being the azimuthal angle around the \(z\)-axis.
The pseudorapidity is defined in terms of the polar angle \(\theta\) as \(\eta = -\ln \tan(\theta/2)\) and is equal to the rapidity
$y = (1/2) \ln[(E + p_z)/(E - p_z)]$ in the relativistic limit.
Angular distance is measured in units of \(\Delta R \equiv \sqrt{(\Delta y)^{2} + (\Delta\phi)^{2}}\).}

The ATLAS detector~\cite{PERF-2007-01} at the LHC covers nearly the entire solid angle around the collision point.\footnote{\AtlasCoordFootnote}
It consists of an inner tracking detector surrounded by a thin superconducting solenoid, electromagnetic and hadronic calorimeters,
and a muon spectrometer incorporating three large superconducting air-core toroidal magnets.

The inner-detector system (ID) is immersed in a \qty{2}{\tesla} axial magnetic field
and provides charged-particle tracking in the range \(|\eta| < 2.5\).
The high-granularity silicon pixel detector covers the vertex region and typically provides four measurements per track,
the first hit generally being in the insertable B-layer (IBL) installed before Run~2~\cite{ATLAS-TDR-19,PIX-2018-001}.
It is followed by a silicon microstrip detector called the SemiConductor Tracker (SCT), which usually provides eight measurements per track, two per double-sided detector layer. The barrel region, $|\eta| < 1.7$, is of special interest to this analysis. Here the pixel detector comprises four layers at radii of 33.25\,mm (IBL), 50.5\,mm, 88.5\,mm, and 122.5\,mm. This is followed by the four SCT barrel layers, located at radii of 299\,mm, 371\,mm, 443\,mm, and 514\,mm. These silicon detectors are complemented by the transition radiation tracker (TRT),
which enables radially extended track reconstruction up to \(|\eta| = 2.0\).
The TRT also provides electron identification information
based on the fraction of hits (typically 30 in total) above a higher energy-deposit threshold corresponding to transition radiation.

The calorimeter system covers the pseudorapidity range \(|\eta| < 4.9\).
Within the region \(|\eta|< 3.2\), electromagnetic calorimetry is provided by barrel and
endcap high-granularity lead/liquid-argon (LAr) calorimeters,
with an additional thin LAr presampler covering \(|\eta| < 1.8\)
to correct for energy loss in material upstream of the calorimeters.
Hadronic calorimetry is provided by the steel/scintillator-tile calorimeter,
segmented into three barrel structures within \(|\eta| < 1.7\), and two copper/LAr hadronic endcap calorimeters.
The solid angle coverage is completed with forward copper/LAr and tungsten/LAr calorimeter modules
optimised for electromagnetic and hadronic energy measurements, respectively.

The muon spectrometer (MS) comprises separate trigger and
high-precision tracking chambers measuring the deflection of muons in a magnetic field generated by the superconducting air-core toroidal magnets.
The field integral of the toroids ranges between \num{2.0} and \qty{6.0}{\tesla\metre}
across most of the detector.
Three layers of precision chambers, each consisting of layers of monitored drift tubes, cover the region \(|\eta| < 2.7\),
complemented by cathode-strip chambers in the forward region, where the background is highest.
The muon trigger system covers the range \(|\eta| < 2.4\) with resistive-plate chambers in the barrel, and thin-gap chambers in the endcap regions.

The luminosity is measured mainly by the LUCID-2 detector~\cite{LUCID2}, which records Cherenkov light produced in the quartz windows of photomultipliers located close to the beam pipe.

Events were selected by the first-level trigger system implemented in custom hardware,
followed by selections made by algorithms implemented in software in the high-level trigger~\cite{TRIG-2016-01}.
The first-level trigger accepted events from the \qty{40}{\MHz} bunch crossings at a rate close to \qty{100}{\kHz},
which the high-level trigger reduced in order to record complete events to disk at about \qty{1.25}{\kHz}.

A software suite~\cite{SOFT-2022-02} is used in data simulation, in the reconstruction
and analysis of real and simulated data, in detector operations, and in the trigger and data acquisition
systems of the experiment.

% End of text imported from the .//sections/detector.tex input file

%
\section{Data and simulated event samples}
\label{sec:DataAndMC}
%

% The next lines are included from the .//sections/data_mc.tex input file
This search uses data from an integrated luminosity of 137~\ifb of \pp collisions at a centre-of-mass energy $\sqs = 13$~\TeV recorded by the ATLAS experiment during Run~2 of the LHC~(2015--2018). In this data-taking period, the number of interactions per bunch crossing ranged from about 8 to 70, with an average of 34~\cite{DAPR-2018-01}. The uncertainty in the integrated luminosity is 0.83\%~\cite{DAPR-2021-01}, obtained using the LUCID-2 detector for the primary luminosity measurements, complemented by measurements using the ID and calorimeters.
Only events recorded under stable beam conditions are used in the analysis, and general data-quality requirements ensure that all detector subsystems were operational~\cite{DAPR-2018-01}. In about 2\% of Run 2 data taking the SCT experienced a significant increase in readout errors compared to standard operation. These data periods are removed as it increases the number of spuriously reconstructed pixel-only tracks which can mimic a disappearing-track signature.

The data were collected primarily with a missing transverse momentum trigger~\cite{ATL-DAQ-PUB-2016-001,TRIG-2019-01}, to target events where the sparticle system recoils against a high-\pt ISR jet and the undetected neutral particles lead to a large amount of missing transverse momentum. The trigger threshold was raised as the luminosity increased during Run~2, but each threshold value provided an efficiency of ${>}90$\% when requiring online a reconstructed missing transverse momentum of at least 230\,GeV. Additional single-lepton ($\ell = e,\mu$) triggers are used to select regions to constrain background  estimates, and these triggers operate on an efficiency plateau when applying an offline lepton transverse momentum requirement of $\pt(\ell) > 27\,\GeV$~\cite{TRIG-2018-01,TRIG-2018-05}.

Monte Carlo (MC) simulated samples are used to model SM processes and estimate the expected signal yields.
All samples were produced using the ATLAS simulation infrastructure~\cite{SOFT-2010-01} and either $\GEANT$~\cite{Agostinelli:2002hh}
or a faster simulation based on a parameterisation of the calorimeter response and \GEANT for the other detector
systems. The effect of multiple interactions in the same and neighbouring bunch crossings (\pileup) was modelled by overlaying each simulated hard-scattering event with inelastic $pp$ events generated with \PYTHIA[8.186]~\cite{Sjostrand:2007gs} using the \NNPDF[2.3lo] set of parton distribution functions (PDFs)~\cite{Ball:2012cx} and the A3 set of tuned parameters~\cite{ATL-PHYS-PUB-2016-017}. The MC simulated samples were reweighted to match the actual distribution of the mean number of interactions per bunch crossing observed in data in 2015--2018.

The electroweakino signal samples were generated with \MGNLO[2.9.9]~\cite{Alwall:2014hca} at leading order (LO) and interfaced to \PYTHIA[8.306]~\cite{Sjostrand:2014zea} for the modelling of the parton showering (PS), hadronisation and underlying event. In the wino scenarios, $\Delta m(\chinoonepm, \ninoone)$ is dependent upon $m(\chinoonepm)$ and follows from the calculations in Refs.~\cite{Giudice:1998xp, Randall:1998uk}. In the higgsino scenarios, the mass splitting is dependent upon $m(\chinoonepm)$ and is set according to the calculations in Ref.~\cite{Nagata:2014wma}. The CMSSM $\tau$-slepton signals were generated with \MGNLO[2.9.5] at LO and interfaced to \PYTHIA[8.306], while the GMSB $\tau$-slepton signals used \MGNLO[2.6.1] at LO and \PYTHIA[8.230].

For all signal scenarios the underlying event was modelled with the A14~\cite{ATL-PHYS-PUB-2014-021} set of tuned parameters. The matrix element (ME) calculation was performed at tree level and includes the emission of up to two additional partons. The ME--PS merging was done using the CKKW-L~\cite{Lonnblad:2012ix} prescription. The \NNPDF[3.0lo] set of PDFs was used throughout~\cite{Ball:2014uwa}. To ensure efficient sample generation, a lifetime reweighting procedure was used, where a single signal-mass scenario and lifetime has a set of event weights allowing the signal to be reweighted to different proper lifetimes of the long-lived particle. The charginos and $\tau$-sleptons are propagated through the detector to their decay point using the standard simulation, including magnetic field effects and electromagnetic interactions with detector material.

Signal cross-sections are calculated to next-to-leading order (NLO) in the strong coupling constant, adding the resummation of soft gluon emission at next-to-leading-logarithm (NLL) accuracy~\cite{Beenakker:1999xh,Debove:2010kf,Fuks:2012qx,Fuks:2013vua,Fiaschi:2018hgm,Bozzi:2007qr,Fuks:2013lya,Fiaschi:2018xdm,newResummino} using \textsc{Resummino}~2.0.1. The nominal cross-section and its uncertainty are taken from an envelope of cross-section predictions using different PDF sets and factorisation and renormalisation scales, as described in Ref.~\cite{Borschensky:2014cia}. For the wino and higgsino signal samples, only processes involving the direct production of a \chinoonepm are considered in order to produce the disappearing-track signature. Therefore, the wino samples include $\chinoonep\chinoonem$ and $\chinoonepm\ninoone$ production, while the higgsino signal samples include $\chinoonep\chinoonem$, $\chinoonepm\ninoone$, and $\chinoonepm\ninotwo$. Taking an example for higgsino production with $m(\chinoonepm) = 300\,\GeV$ and $m(\ninotwo) = m(\ninoone) = 299\,\GeV$, the cross-section value for each of the $\chinoonepm\ninotwo$ and $\chinoonepm\ninoone$ processes is $90.8$\,fb, while the cross-section is $53$\,fb for $\chinoonep\chinoonem$ production.
For wino production with $m(\chinoonepm) = 300\,\GeV$ and $m(\ninoone) = 299\,\GeV$,
the cross-section values are $387$\,fb and $190$\,fb for $\ninoone\chinoonepm$ and $\chinoonep\chinoonem$ production, respectively.
Finally, the cross-section for $\tau$-slepton pair production of $m(\stau)= 300\,\GeV$ is $6.3$\,fb in both the CMSSM and GMSB models.

MC simulation of $V$+\;\!jets events was used to study the fake-tracklet background. They were simulated with the \SHERPA[2.2.1]~\cite{Bothmann:2019yzt} generator using NLO matrix elements for up to two partons, and LO matrix elements for up to four partons, calculated with the Comix~\cite{Gleisberg:2008fv} and \OPENLOOPS~\cite{Buccioni:2019sur,Cascioli:2011va,Denner:2016kdg} libraries. The matrix elements were matched with the \SHERPA parton shower~\cite{Schumann:2007mg} using the \MEPSatNLO prescription~\cite{Hoeche:2011fd,Hoeche:2012yf,Catani:2001cc,Hoeche:2009rj} and the set of tuned parameters developed by the \SHERPA authors. The \NNPDF[3.0nnlo] set of PDFs~\cite{Ball:2014uwa} was used and the samples were normalised to a NNLO prediction~\cite{Anastasiou:2003ds}.

To evaluate the detector response when reconstructing short tracks and to calculate the associated reconstruction efficiencies, a set of so-called \enquote{particle gun} samples are used. These samples do not represent a specific physical event or process; instead, for each generated event the particle of interest is placed at the centre of the detector with a certain momentum and direction and can be used to investigate detector reconstruction capabilities. These samples do not include \pileup. The analysis uses three \enquote{particle gun} samples, corresponding to electrons, muons and charged pions (implicitly including the decay products of $\tau$-leptons), generated uniformly across a  transverse momentum range of $10{-}100\,\GeV$; they are used to estimate the reconstruction efficiencies as needed for the data-driven background estimation methods employed.

% End of text imported from the .//sections/data_mc.tex input file

%
\section{Event reconstruction}
\label{sec:Reco}
%

% The next lines are included from the .//sections/reco.tex input file
The analysis uses standard ATLAS reconstruction techniques to reconstruct leptons and jets, and to calculate the missing transverse momentum, enabling the identification of potential events of interest before employing a dedicated tracklet algorithm to specifically identify events containing disappearing tracks.

Standard tracks are reconstructed in the ID within $|\eta| < 2.5$~\cite{PERF-2015-08} and subsequently combined to create primary vertex candidates formed from at least two tracks with $\pt> 500\,\MeV$~\cite{ATL-PHYS-PUB-2015-026}. Among these, the primary vertex is identified as the vertex with the largest $\sum{\pt^2}$, where the sum runs over tracks associated with the vertex. All events considered in the analysis are required to have a primary vertex. The transverse and longitudinal impact parameters of all tracks, denoted by $d_0$ and $z_0$ respectively, are calculated relative to the beamline. The $z$-position of the primary vertex is used to calculate the point of closest approach, which is used as the reference point to calculate $z_0$.

Electron candidates are reconstructed from clustered energy deposits in the electromagnetic calorimeter matched to tracks re-fitted to account for bremsstrahlung losses~\cite{EGAM-2018-01}. They are required to have $\pt > 10\,\GeV$ and $|\eta| < 2.47$, and must satisfy the \enquote{LooseAndBLayer} quality criteria~\cite{PERF-2017-01}. In order to ensure that the trajectories of the electrons are consistent with originating from the primary vertex, the longitudinal impact parameter must satisfy $|z_0\sin\theta| < 0.5\,$mm, while the significance of the transverse impact parameter, $|d_0|/\sigma(d_0)$, must be less than 5. Electrons are required to be isolated from other objects by assessing a combination of track- and calorimeter-based information at the \enquote{FCLoose} working point~\cite{EGAM-2021-01}.

Muon candidates are reconstructed by combining ID tracks with matching MS tracks. They are required to have $\pt > 10$~\GeV and $|\eta| < 2.7$, and must satisfy the \enquote{Medium} quality criteria~\cite{MUON-2018-03}. Muons originating from the primary vertex are selected by requiring $|z_0\sin\theta| < 0.5\,$mm and $|d_0|/\sigma(d_0) < 3$. The muons must satisfy isolation requirements similar to those applied to electrons, but using the \enquote{Loose\_FixedRad} working point~\cite{MUON-2022-01}. Muons reconstructed purely from MS information are used to estimate muon backgrounds, and are referred to as stand-alone muons.

Hadronic jets are reconstructed using the \antikt algorithm~\cite{Cacciari:2008gp} as implemented in \texttt{FastJet}~\cite{Fastjet} with a jet radius parameter $R = 0.4$. The input to this algorithm consists of particle flow~(PFlow) objects~\cite{PERF-2015-09}, which combine measurements from the ID and calorimeters~\cite{PERF-2014-07} to improve the jet energy resolution and increase the jet reconstruction efficiency, especially at low jet $\pT$. Calibrations are applied to the jet energy scale and resolution, and these include components derived from both simulation and in situ measurements~\cite{JETM-2018-05}. The analysis uses jets within $|\eta| < 2.8$ with a calibrated $\pt > 20\,\GeV$. In order to reduce contributions from jets produced in \pileup interactions, all jets with $|\eta| < 2.5$ and $\pt < 60\,\GeV$ are required to satisfy the jet-to-vertex tagger~(JVT)~\cite{PERF-2014-03} \enquote{Tight} working point requirements. The tagger is configured to identify jets from the hard-scatter vertex with 92\% efficiency.

An overlap removal procedure is applied to avoid double counting of energy from different types of objects. If an electron and a jet are reconstructed within $\Delta R < 0.2$, the jet is rejected, while the electron candidate is kept. Subsequently, if an electron or muon is separated from a jet by $\Delta R < 0.4$, the electron or muon is discarded and the jet is kept.

The missing transverse momentum, with magnitude \met, is calculated as the negative vector sum of the transverse momenta of all identified hard physics objects (electrons, muons, jets), plus a contribution from an additional \enquote{soft term}. Forward jets within $2.8 < |\eta| < 4.5$ are also included in the \met calculation. The soft term is calculated from tracks matched to the primary vertex and not assigned to any of the hard objects~\cite{JETM-2020-03} but does not contain the reconstructed tracklets or charged pions described in the following paragraphs.

To reconstruct the disappearing track due to a charged particle decaying after passing through the pixel detector, a dedicated tracklet reconstruction algorithm is applied after the standard reconstruction techniques described above are performed~\cite{ATL-PHYS-PUB-2019-011}. The tracking-detector hits already associated with reconstructed standard tracks are excluded from this tracklet-finding iteration. The algorithm attempts to identify tracklets having three or four hits in the pixel detector. The procedure attempts to extend the tracks into the SCT and TRT detectors and vetoes the tracklet if it can be associated with additional hits. Compared to the previous ATLAS analysis using the disappearing-track signature~\cite{SUSY-2018-19}, the veto does not consider an association if more SCT hits are only found multiple layers away from the tracklet. This reduces cases where tracklets would have been vetoed due to spurious SCT hits~\cite{ATL-PHYS-PUB-2019-011} further into the detector. To improve the momentum resolution of the tracklets, they are re-fitted, adding the primary vertex as an additional constraint, and the tracklet is required to have $|d_0|/\sigma(d_0) < 1.5$ and $|z_0\sin\theta|< 0.2$\,mm. An overlap removal procedure discards tracklets within $\Delta R = 0.4$ of a jet, a lepton, or a track in the MS. Isolation requirements are also applied to the tracklets: the scalar sum of the \pt of all tracks within a cone of radius $\Delta R=0.4$ around the tracklet, $p_{\mathrm{T}}^{\mathrm{cone40}}$, must satisfy $p_{\mathrm{T}}^{\mathrm{cone40}}/p_{\mathrm{T}}^{\mathrm{tracklet}} < 0.04$ and the total transverse energy of all topological clusters within the $\Delta R=0.4$ cone, $E_{\mathrm{T}}^{\mathrm{topoclus40}}$, must be less than 5\,\GeV. Tracklets are required to have $|\eta| > 0.1$, vetoing the central region where MS coverage is reduced. Figure~\ref{fig:reco:TrackEff} shows that the reconstruction efficiency for three- and four-layer tracklets can reach 80\% depending upon the distance travelled by the chargino before decaying.

To identify low-energy charged pions, indicative of the electroweakino signal scenario, a boosted decision tree (BDT) machine-learning algorithm is used. After performing the tracklet reconstruction described above, a set of low-momentum tracks (\enquote{soft tracks}) are reconstructed from the remaining hits. These low-momentum tracks are seeded from at least three SCT hits that are within $\Delta R = 0.8$ of a tracklet and are then extended using further SCT or TRT hits. A minimum of six hits, and at most two holes\footnote{A hole is a missing hit on a reconstructed track trajectory, possibly due to a non-functioning strip which was not masked.}~\cite{ATL-PHYS-PUB-2019-011}, are required to reconstruct a soft track and a requirement of $\pt > 200$\,MeV is applied. As only a limited number of four-layer tracklet events have a matching soft-track that satisfies the soft-track reconstruction requirements, only events containing tracklets produced with three pixel hits are used when performing the pion reconstruction. The BDT is used to select viable charged pion candidates and to reject events with coincidental hits forming a soft track. The algorithm is trained with MC signal simulation, using soft tracks that have been \enquote{truth-matched} at generator level to charged pions from long-lived chargino decays using a $\Delta R$-matching criterion.

The background is taken from data, selecting tracklets that have large impact parameter values with respect to the primary vertex. From MC studies this is expected to be dominated by accidental crossings, referred to as the \enquote{fake track} background. The algorithm uses eight input variables related to the properties of the soft tracks, including the kinematic properties \pt and $\eta$, the number of hits/holes associated with the track in the SCT, the number of hits associated with the track in the TRT, the distance in $\Delta R$ between the soft track and the tracklet, the distance in $z$ between the soft track and the tracklet, and the $\chi^2$ value of the soft track when fitted to the tracklet's endpoint. The TMVA architecture~\cite{TMVA} is used with gradient boost, and a hyperparameter optimisation is performed to prevent overtraining. Figure~\ref{fig:reco:pionBDT} displays the BDT output score. A low-energy charged pion is identified if the score for a soft track is ${>}0.8$ and, to reduce the misidentification rate, a transverse impact parameter requirement of $|d_0|/\sigma(d_0) < 40$ is satisfied. The latter rejects an additional 4\% of the tracklet background with minimal impact on the signal, according to MC simulation. Figure~\ref{fig:reco:pionEff} shows that the pion reconstruction efficiency, measured as a function of the true pion \pt in the lab-frame, is as high as 70\% for \pt near 2\,\GeV.

\begin{figure}[tb!]
\begin{center}
\subfigure[]{
\includegraphics[width=0.44\textwidth]{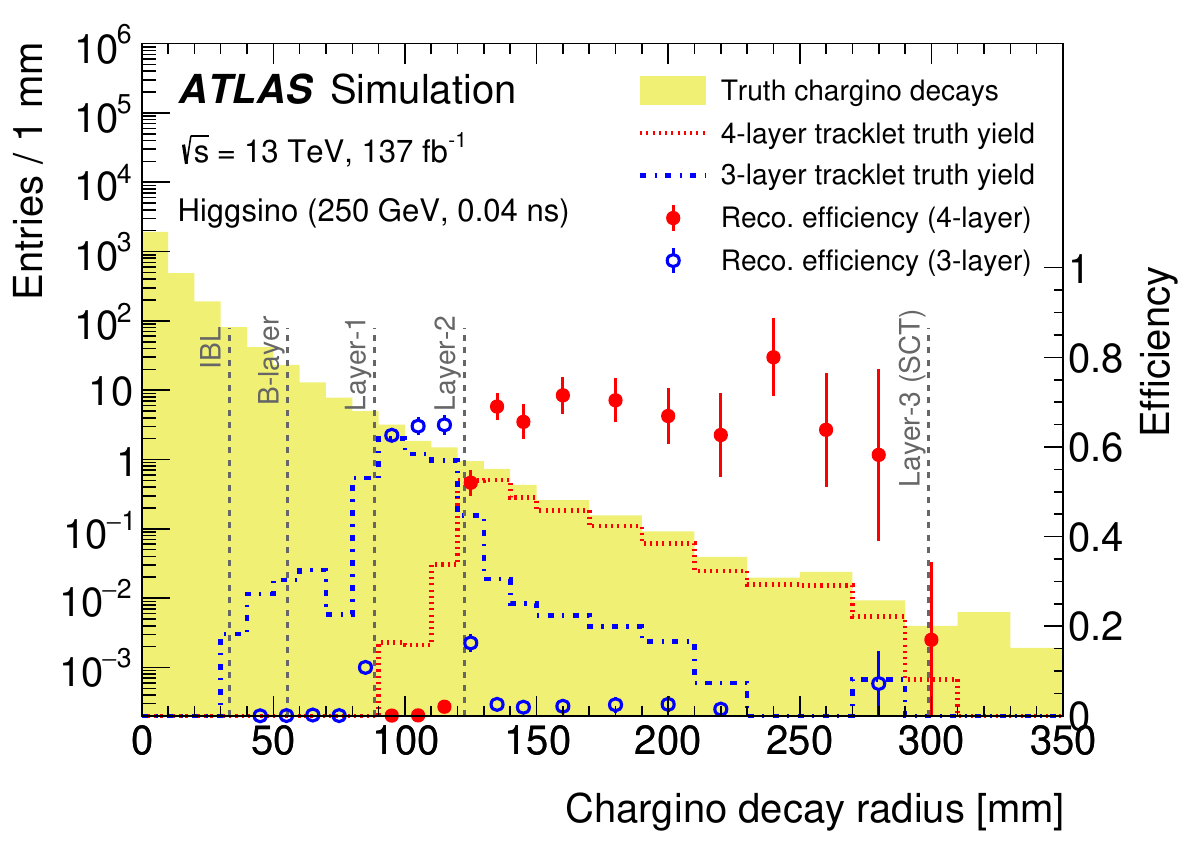}
\label{fig:reco:TrackEff}
}
\subfigure[]{
\includegraphics[width=0.44\textwidth]{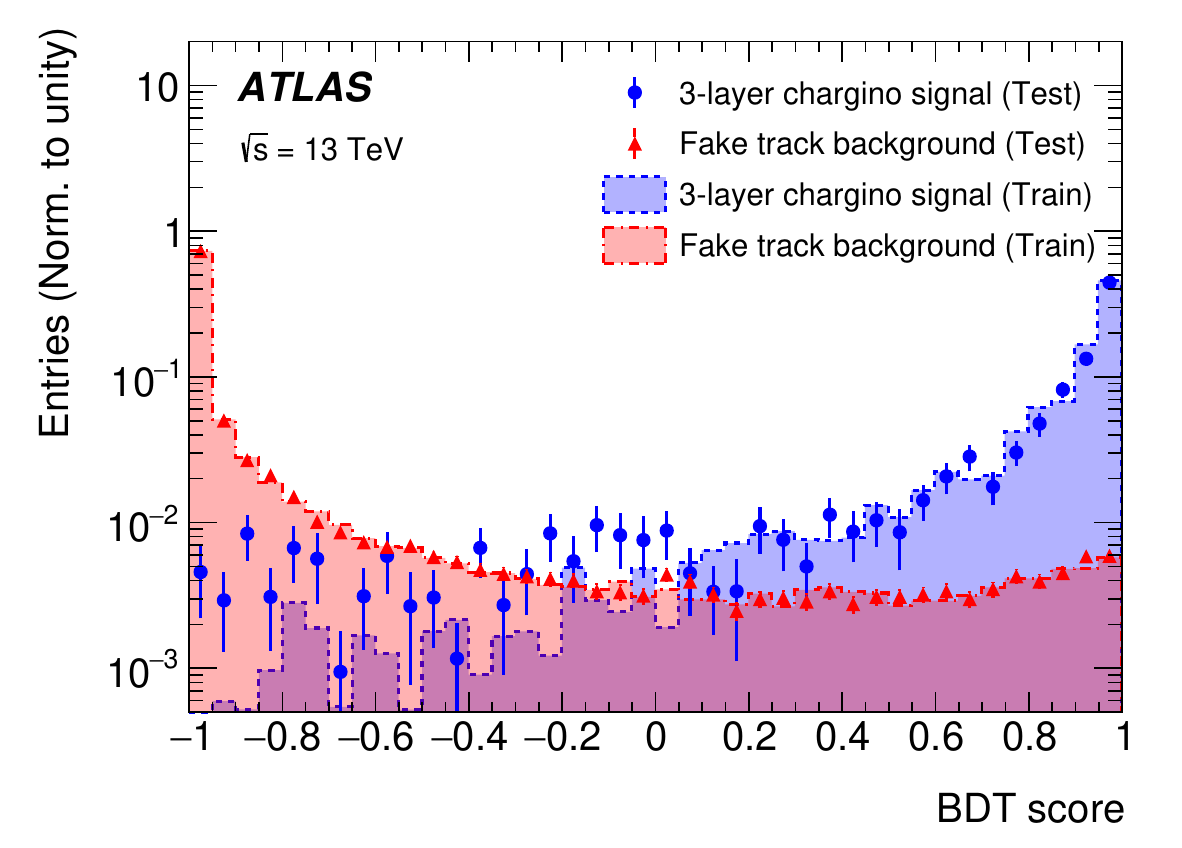}
\label{fig:reco:pionBDT}
}
\subfigure[]{
\includegraphics[width=0.44\textwidth]{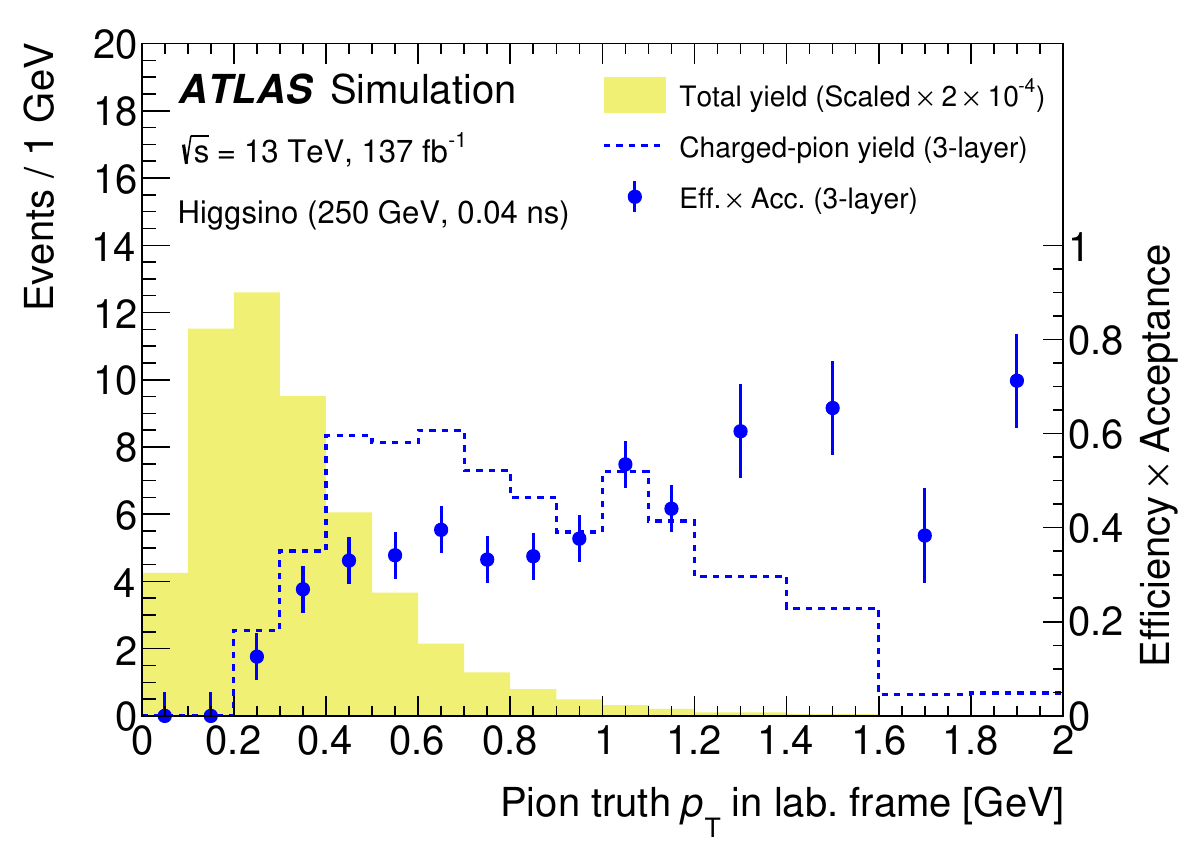}
\label{fig:reco:pionEff}
}
\end{center}
\vspace{-0.75cm}
\caption{Representative distributions related to the reconstruction of tracklets and charged pions: \protect\subref{fig:reco:TrackEff} event yields and reconstruction efficiencies for  three- and four-layer tracklet reconstruction as a function of \chinoonepm decay radius; \protect\subref{fig:reco:pionBDT} BDT output score used to judge if an event contains a low-energy pion, showing the training dataset (solid histogram) and the test dataset (points); \protect\subref{fig:reco:pionEff} event yields and reconstruction efficiency for charged pions as a function of the true pion transverse momentum as measured in the lab-frame, where the acceptance includes all charged-pion reconstruction effects but not geometric acceptance from tracklet reconstruction (which is assumed to already have been reconstructed). The true transverse momentum is defined as $\pt = \sqrt{p_x^2 + p_y^2}$ in the global detector coordinate system with the $z$-axis along the beam pipe. For \protect\subref{fig:reco:TrackEff} and \protect\subref{fig:reco:pionEff} the filled histogram displays the total yield without applying a selection, while the dashed lines present the yield after applying the tracklet selection or tracklet+pion selection. The points denote the reconstruction efficiency. A representative higgsino signal scenario with $m(\chinoonepm) = 250$\,GeV and $\tau(\chinoonepm)= 0.04$\,ns is shown.
}
\end{figure}

\FloatBarrier

% End of text imported from the .//sections/reco.tex input file

%
\section{Analysis strategy}
\label{sec:Analysis}
%

% The next lines are included from the .//sections/analysis.tex input file
The signal scenarios investigated contain a tracklet, \met, and in the case of the chargino production scenarios a low-energy charged pion. In all considered signal scenarios, the \met is however too small to satisfy the \met trigger requirements. Therefore, events are required to contain a high-\pt initial-state radiation (ISR) jet to boost the SUSY particle system and achieve enough \met to exceed the trigger threshold. Signal regions (SRs) are defined using kinematic variable selections to enhance the contribution from signal processes relative to the background processes. All SRs require events to have zero leptons, at least one jet, and a large azimuthal separation between the highest-\pt jet and the $\vec{p}_{\mathrm{T}}^{\mathrm{miss}}$, denoted by $|\Delta\phi(j_1,\vec{p}_{\mathrm{T}}^{\mathrm{miss}})|$. Four orthogonal SRs are defined in the analysis, two requiring the presence of a tracklet reconstructed from four pixel layers, and another two requiring the presence of a tracklet reconstructed from three pixel layers and differing mainly in whether or not a low-energy charged pion is required. Table~\ref{tab:SRA_Selections} provides a summary of all SR selections used in the analysis.

\begin{table}[tb!]
\centering
\caption{Summary of the signal region selection requirements. \label{tab:SRA_Selections}}
\begin{tabular}{l | c | c | c | c }
\toprule
Variable & \SRAh & \SRAm & \SRBo & \SRBz \\
\midrule
\met trigger & \multicolumn{4}{c}{Passed} \\
Number of leptons &   \multicolumn{4}{c}{$= 0$} \\
Number of jets ($\pt > 20\,\GeV$) &   \multicolumn{4}{c}{$\geq 1$} \\
Leading jet \pt &   \multicolumn{4}{c}{$> 100\,\GeV$} \\
$|\Delta\phi(j_1,\vec{p}_{\mathrm{T}}^{\mathrm{miss}})|$ &   \multicolumn{4}{c}{$> 1.0$} \\
Number of four-layer tracklets &   \multicolumn{2}{c|}{$\geq 1$} & \multicolumn{2}{c}{$= 0$} \\
Number of three-layer tracklets &   \multicolumn{2}{c|}{-} & \multicolumn{2}{c}{$\geq 1$} \\
Number of charged pions & \multicolumn{2}{c|}{-} & $\geq 1$ & $= 0$ \\
Tracklet $p_{\mathrm{T}}$ & $> 60\,\GeV$ & $> 60$\,GeV & $> 60\,\GeV$ & $> 60\,\GeV$ \\
Tracklet $|\eta|$ & 0.1--1.9 & 0.1--1.9 & 0.1--1.9 & 0.1--1.7 \\
\met & $> 300$\,GeV & $150{-}300\,\GeV$ & $>240\,\GeV$ & $>280\,\GeV$ \\
\bottomrule
\end{tabular}
\end{table}

The four-hit SRs (\SRAh and \SRAm) require events to contain a tracklet that is reconstructed from four pixel-detector hits and has tracklet $p_{\mathrm{T}} >60\,\GeV$. Tracklets with hits in four layers provide sensitivity to a wide range of \chinopm/$\tau$-slepton lifetimes, from 0.02 to 10\,ns. No requirement is placed on the presence of a low-energy charged pion. \SRAh requires high \met, whereas \SRAm selects events with lower \met. The two SRs containing tracklets reconstructed from hits in three layers are defined in order to target very short-lived charged SUSY particles, which is the phase space of interest for the higgsino-like case, where $\tau(\chinoonepm) \sim 0.02{-}0.2$\,ns. A tracklet $p_{\mathrm{T}} > 60\,\GeV$ requirement is applied for both three-layer tracklet SRs and the main difference between the two is the absence or presence of a low-energy charged pion. \SRBo requires the presence of at least one low-energy charged pion and requires $\met >240\,\GeV$, whereas \SRBz applies a low-energy charged-pion veto and applies a slightly tighter $\met > 280\,\GeV$ selection.

In all regions, the backgrounds arise either from events containing particles that are misidentified as tracklets or from unrelated coincidental pixel-layer hits that form a tracklet. Figure~\ref{fig:analysis:BkgSchematic} presents schematically how background events might be mis-reconstructed and satisfy the SR selections. The \chinoonepm signal scenario is also shown, in the case where the charged pion is reconstructed from the track it leaves in the SCT detector, and in the case where the charged pion does not have enough energy to reach the SCT and leave any hits. Background events may have tracklets reconstructed from pixel hits which belong to known SM particles (electrons, muons, hadrons) that do not pass standard ATLAS reconstruction techniques or from coincidentally combined hits (referred to as the \enquote{fake} background).

\begin{figure}[t!]
\centering
\includegraphics[width=0.95\textwidth]{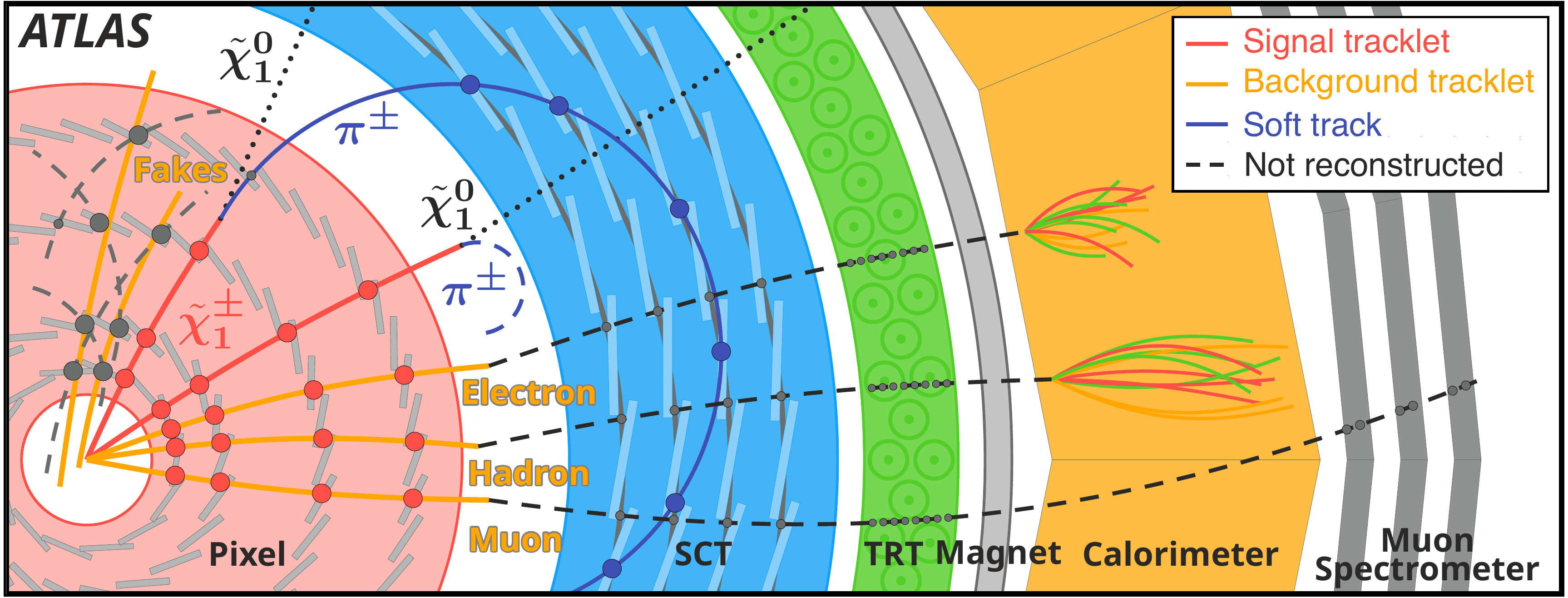}
\caption{Schematic diagram illustrating how the signal scenarios are detected and how the background processes may enter the SR selections. In the case of SM particles, the small grey circles indicate measurements that are not associated with the electron/hadron/muon in standard reconstruction, allowing the SM particles to enter the SR selections. The detector layers are not shown to scale. Taken from Ref.~\cite{SUSY-2018-19}.}
\label{fig:analysis:BkgSchematic}
\end{figure}

\FloatBarrier

% End of text imported from the .//sections/analysis.tex input file

%
\section{Background estimation}
\label{sec:bkgs}
%

% The next lines are included from the .//sections/backgrounds.tex input file
Data-driven methods are used to estimate all backgrounds in the analysis and a similar approach is used for all background sources. Template regions (TRs) are used to calculate an initial tracklet \pt distribution/template by selecting events containing the background under consideration. A set of alternative regions, with relaxed reconstruction selections, are used to calculate transfer factors (TFs) that are then applied to the tracklet \pt template to correct for the rate of electrons, muons, hadrons, and spurious signals/fakes misidentified as tracklets. This approach produces a purely data-driven estimate of the SR background from electrons and muons. However, MC simulation is used to calculate the TFs for the hadronic and fake backgrounds, so control regions (CRs) are used to normalise the tracklet \pt distribution and predict the expected yields in the SRs.

The electron and muon backgrounds arise due to bremsstrahlung from interactions with the inner-detector material resulting in an abrupt increase in the electron/muon track curvature. This causes the lepton to not satisfy the standard tracking requirements for lepton reconstruction and it is instead incorrectly reconstructed as a tracklet. Hadrons can also be misidentified as tracklets because of similar sudden increases in track curvature due to bremsstrahlung or hadronic interactions in the inner-detector material, or charged hadron decays inside the inner detector. Fake tracklets can be formed with spurious pixel hits from low-\pt particles that do not belong to the primary interaction (\pileup), from the underlying event, or from detector noise. The methods described below are used to estimate the expected tracklet \pt distributions ($f_{\mathrm{SR}}^{\mathrm{bkg}}$) in each SR.

The dominant background contribution in all SRs is due to fake tracklets. From MC studies it is found that true tracklets usually exhibit a correlation between the true particle \pt and the tracklet \pt, whereas fake tracklets do not have this correlation. For each SR the \pt distribution of fake tracklets is estimated by

\begin{equation*}
f_{\mathrm{SR}}^{\mathrm{fake}}(\pt) = N^{\mathrm{TRfake}}(\pt) \times \mathrm{TF}^{\mathrm{fake}}_{d_0\textrm{-correction}}(d_0) \times \mathrm{TF}^{\mathrm{fake}}_{\mathrm{hybrid}}(\pt) \times \mathrm{SF}^{\mathrm{fake}}_{z_0} \times \mathrm{SF}^{\mathrm{fake}}_{\met},
\end{equation*}

using the yields in TRs modified by TFs and scale factors (SFs), which are described further below. In total, three tracklet \pt templates ($N^{\mathrm{TRfake}}(\pt)$) are extracted from three TRs (TRfake), each defined with the same selection as the SRs, but with orthogonal selections of $\met < 150\,\GeV$ and $|z_0\sin\theta| > 2.5$\,mm. Due to this inverted \met selection, the same TRfake is used for \SRAh and \SRAm. The \met trigger can be safely used with this lower selection because the trigger efficiency does not need to be well understood in this region since it is not used to predict the overall event yield. Furthermore, because the tracklets are not included in the \met calculation, the shape of the \pt distribution is independent of \met. The inverted impact parameter selection is highly efficient and results in a negligible amount of signal entering the region. All three TRs select tracklets reconstructed from hits in three layers to ensure a high-statistics \pt template. Figure~\ref{fig:bkg:34_layer_Template} presents an example of the $N^{\mathrm{TRfake}}(\pt)$ distribution extracted from the TR associated with \SRAh and \SRAm.

From studies using $V$+\:\!jets MC events, it is found that extrapolating from the high-$|z_0\sin\theta|$ TR to regions with low $|z_0\sin\theta|$ introduces a $|d_0|/\sigma(d_0)$ dependence. This is corrected for by using $|d_0|/\sigma(d_0)$-binned correction factors, $\mathrm{TF}^{\mathrm{fake}}_{d_0\textrm{-correction}}$, defined as the ratio of the $|d_0|/\sigma(d_0)$ distribution shape for tracklets with $|z_0\sin\theta| < 0.2\,$mm to that for those with $|z_0\sin\theta| > 2.5\,$mm and are taken directly from data. Figure~\ref{fig:bkg:fake_SF_d0} shows the difference in $|d_0|/\sigma(d_0)$ between the TR and a low-$|z_0\sin\theta|$ region, with the $\mathrm{TF}^{\mathrm{fake}}_{d_0\textrm{-correction}}$ shown in the lower panel.

\begin{figure}[tb!]
\begin{center}
\subfigure[]{
\includegraphics[width=0.44\textwidth]{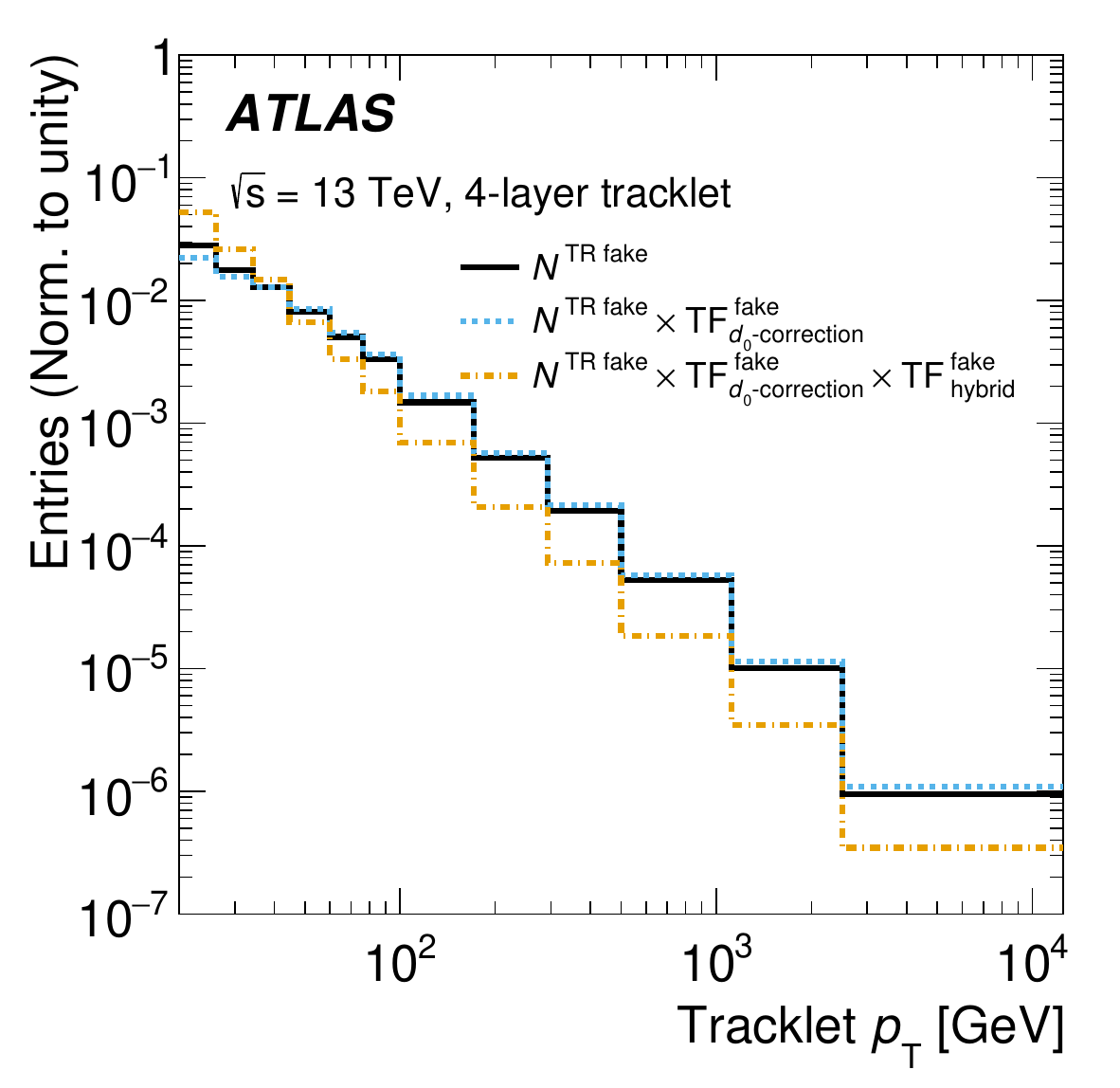}
\label{fig:bkg:34_layer_Template}
}
\subfigure[]{
\includegraphics[width=0.44\textwidth]{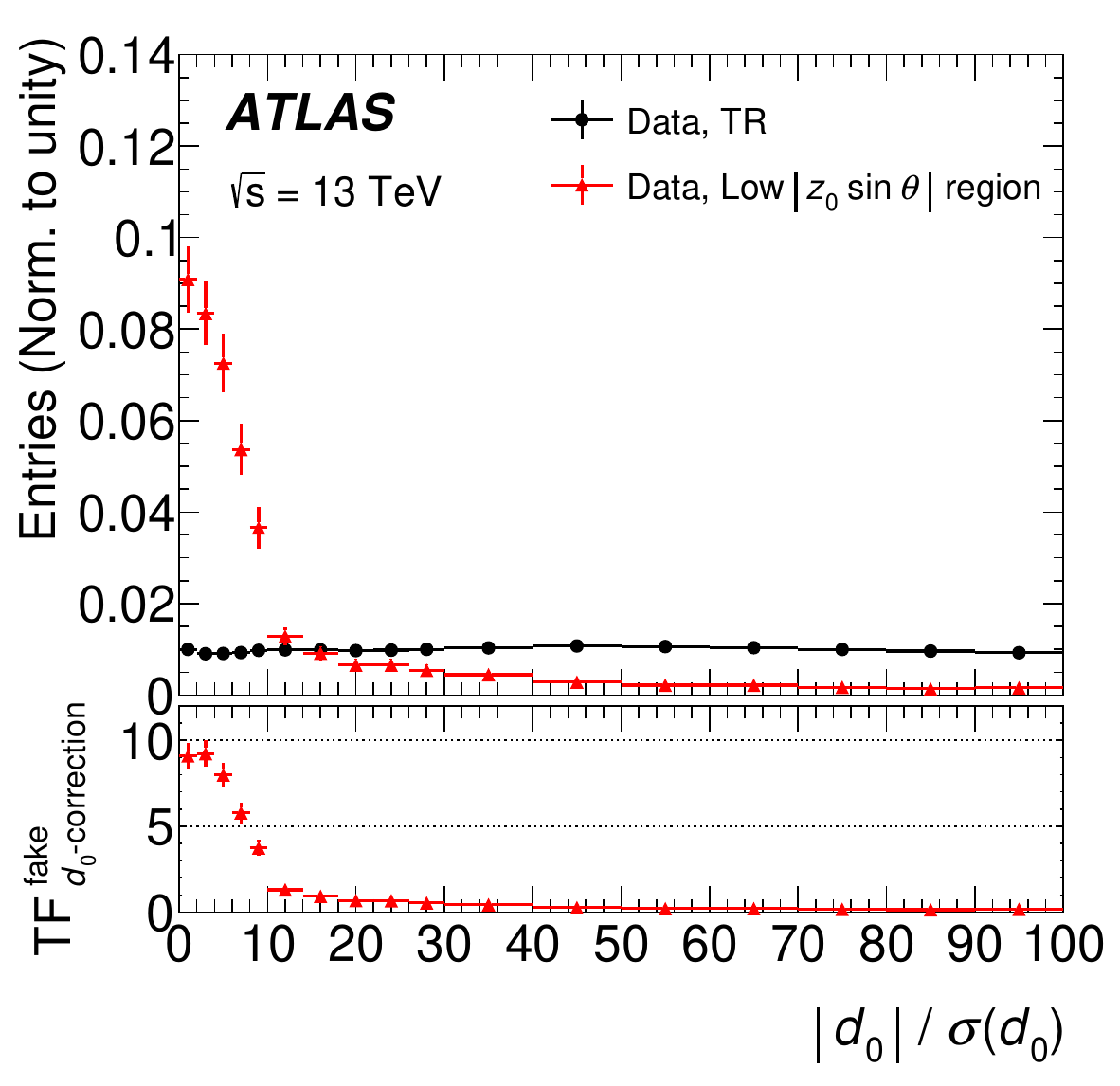}
\label{fig:bkg:fake_SF_d0}
}
\subfigure[]{
\includegraphics[width=0.44\textwidth]{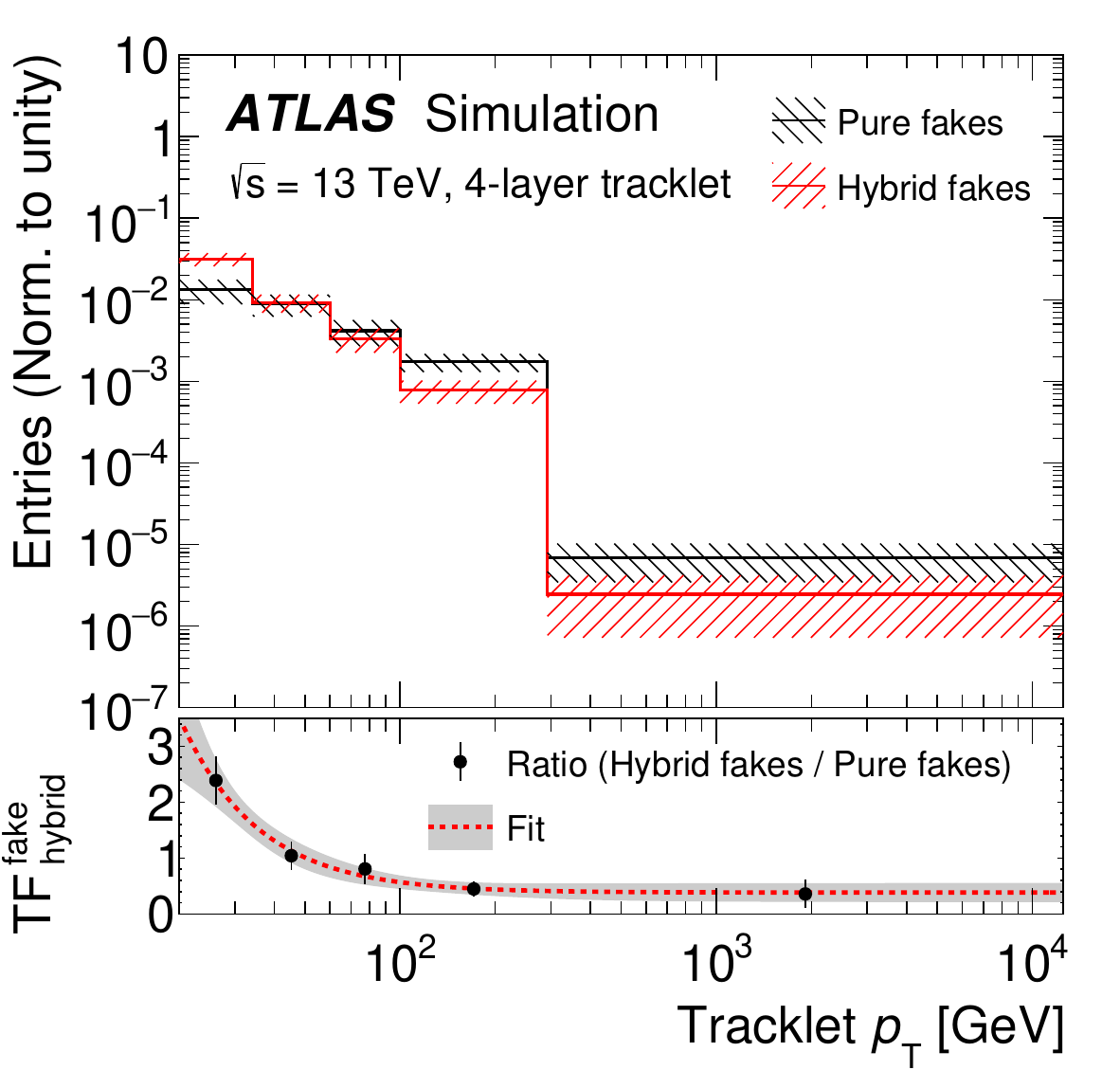}
\label{fig:bkg:34_Hybrid}
}
\end{center}
\vspace{-0.75cm}
\caption{Individual components used in the estimation of the fake background: \protect\subref{fig:bkg:34_layer_Template} shows the original template, $N^{\mathrm{TRfake}}(\pt)$, as a solid line. The templates after the application of the transfer factors are shown as dashed lines; \protect\subref{fig:bkg:fake_SF_d0} shows the $|d_{0}|/\sigma(d_{0})$ dependence introduced when extrapolating over $|z_0\sin\theta|$ and the binwise, with respect to $|d_{0}|/\sigma(d_{0})$, transfer factors $\mathrm{TF}^{\mathrm{fake}}_{d_0\textrm{-correction}}(d_0)$ used to remove the dependence; and \protect\subref{fig:bkg:34_Hybrid} shows a comparison between the pure fake and hybrid fake contributions and the \pt-dependent transfer factors $\mathrm{TF}^{\mathrm{fake}}_{\mathrm{hybrid}}(\pt)$ used to extrapolate from the template region that contains pure fakes to the SR, which contains a mixture of hybrid and pure fake components. }
\end{figure}

Studies of $V$+\;\!jets MC events show that the fake-tracklet background has two sources: \enquote{pure} fakes, where the tracklet is reconstructed from unrelated coincidental hits in the pixel layers, and \enquote{hybrid} fakes, where at least one hit is not from a true hadron. In these studies it is found that at low \met the dominant component comes from pure fakes, while at higher \met there is a mixture of both contributions. To account for the different contributions when extrapolating from low- to high-\met regions, the tracklet \pt template is adjusted by a \pt-dependent transfer factor, $\mathrm{TF}_{\mathrm{hybrid}}^{\mathrm{fake}}$, derived from $V$+\:\!jets MC samples and calculated as the ratio of the tracklet \pt distributions for pure fakes and hybrid fakes. The TFs are used to scale the pure fake \pt distribution to the hybrid fake distribution, and the sum of the pure fakes and hybrid fakes is then used for the final tracklet \pt shape. Figure~\ref{fig:bkg:34_Hybrid} shows the relative contributions of pure fakes and hybrid fakes from MC simulation studies, and the ratio of the two is fitted with a falling exponential-plus-constant distribution as shown in the bottom panel.

To scale the template to data, a set of regions enriched in fake tracklets are used; these select events with a low-\pt tracklet ($\pt< 35\,\GeV$), low \met ($\met< 150\,\GeV$), and a high-\pt \enquote{unconstrained} tracklet ($\pt> 100\,\GeV$), while keeping all other SR selections the same. The unconstrained tracklet is produced from the same hits as a typical tracklet, but there is no requirement to re-fit this tracklet to the primary vertex, so a significantly different \pt measurement can result. A set of scale factors, $\mathrm{SF}_{z_0}^{\mathrm{fake}}$, are calculated as the ratio of events in the fake tracklet enhanced regions to the events in the associated TRfake region but with an alternative selection of tracklet $\pt < 35\,\GeV$ and unconstrained-tracklet $\pt <100\,\GeV$ ensuring orthogonality.

A final set of scale factors, $\mathrm{SF}_{\met}^{\mathrm{fake}}$, are used to extrapolate the template from the low-\met region to the SRs with tighter \met selections. The scale factors are derived from CRs which use high values of the tracklet $|d_{0}|/\sigma(d_{0})$ (${>}1.5$), a loosened tracklet \pt selection (${>}20\,\GeV$), but otherwise the same selections as the SRs. MC simulation shows the normalisation regions are ${>}99$\% pure in fake tracklets, with a ratio of pure to hybrid fakes similar to that in the SRs. Table~\ref{tab:CRA_Selections} defines the control regions \CRAl and \CRAh, with the ratio of events in the low-CR to the high-CR used to derive the fake background $\mathrm{SF}_{\met}^{\mathrm{fake}}$ in the signal regions \SRAh and \SRAm. Table~\ref{tab:CR1p_Selections} defines the control regions \CRBol and \CRBoh used to calculate $\mathrm{SF}_{\met}^{\mathrm{fake}}$ for the fake background in \SRBo, and Table~\ref{tab:CR1z_Selections} defines \CRBzl and \CRBzh, which are used to scale this background in \SRBz. The remaining two CRs, \CRBom and \CRBzm, are used to calculate $\mathrm{SF}_{\met}^{\mathrm{fake}}$ for the validation regions, introduced later in the section.

The background from electrons reconstructed as tracklets is given by

\begin{equation*}
f_{\mathrm{SR}}^{e}(\pt,\eta) = N^{\mathrm{TR}e}(\pt,\eta) \times \mathrm{TF}^e_{\mathrm{tracklet}}(\pt,\eta) \times \mathrm{TF}^e_{\textrm{calo-veto}}(\pt,\eta) \times f^{e}_{\mathrm{smear}}(\pt),
\end{equation*}

\begin{figure}[b!]
\begin{center}
\subfigure[]{
\includegraphics[width=0.45\textwidth]{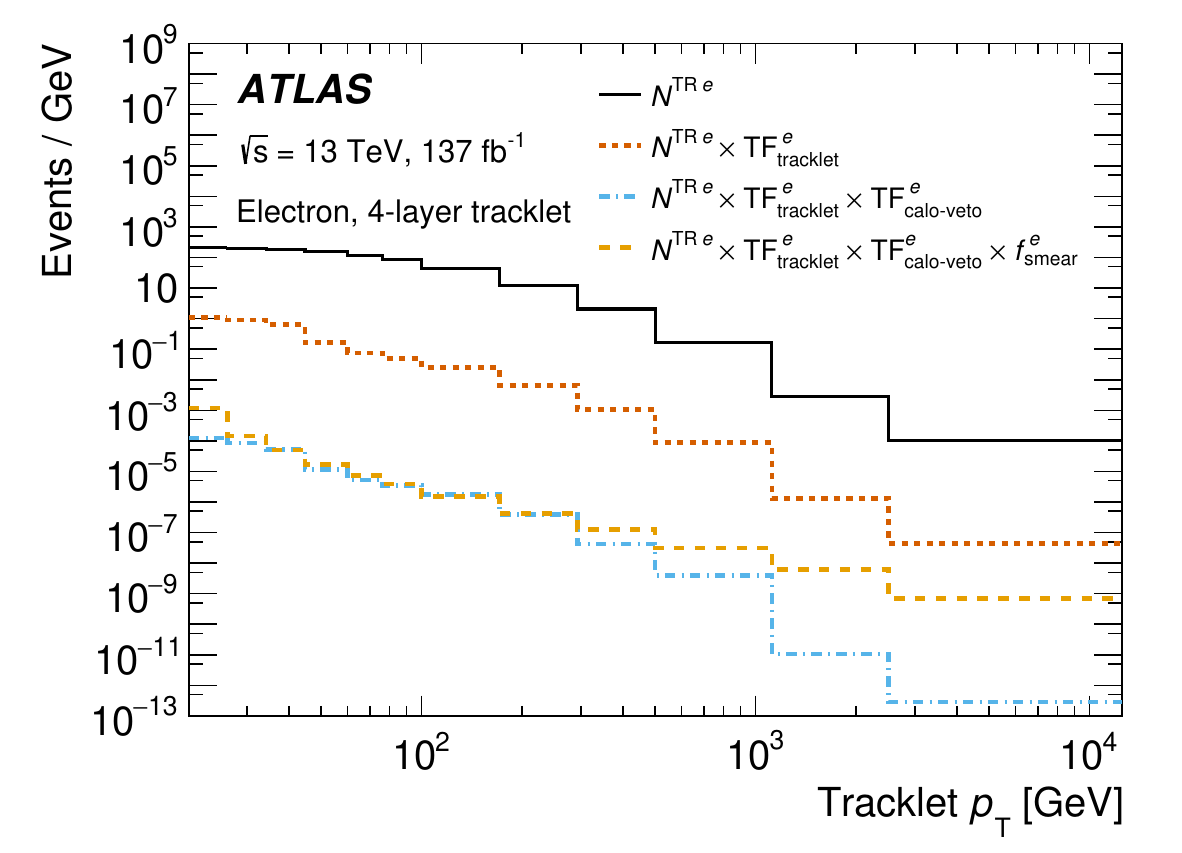}
\label{fig:bkg:el_Template}
}
\subfigure[]{
\includegraphics[width=0.45\textwidth]{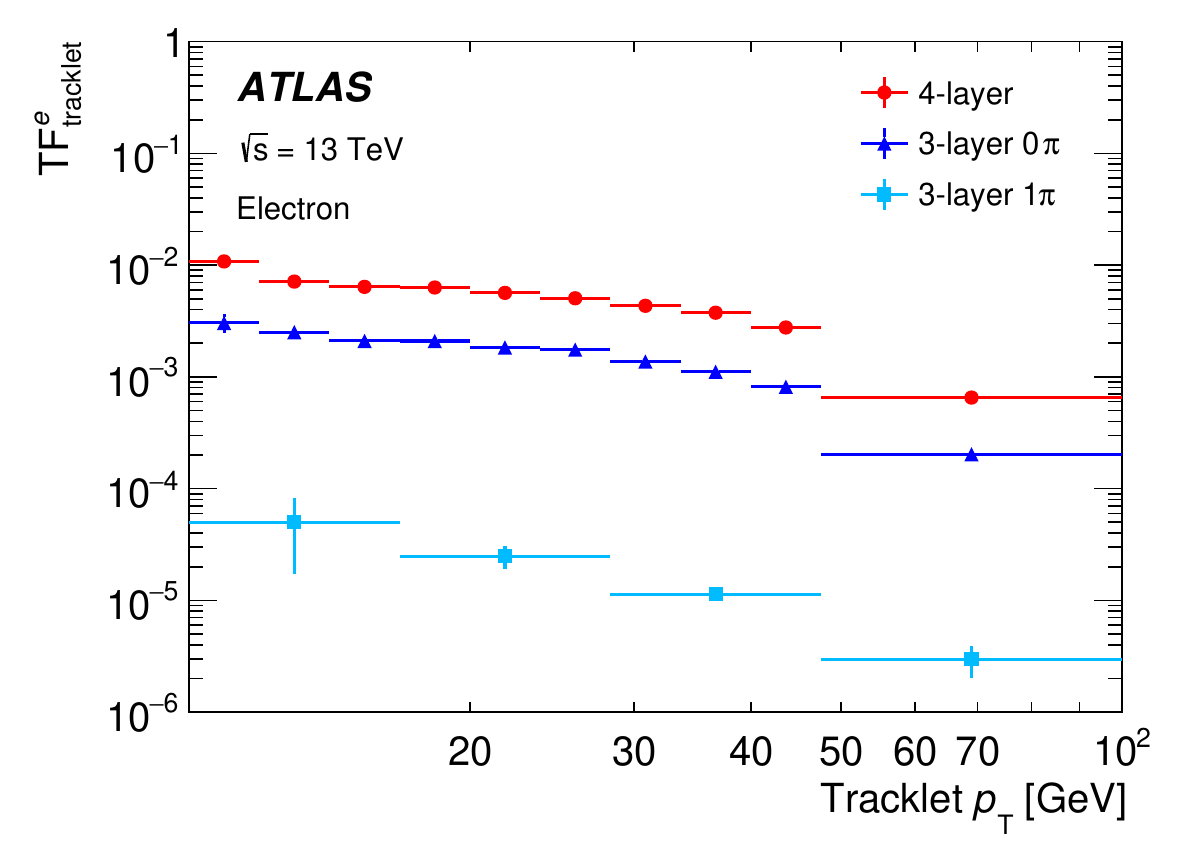}
\label{fig:bkg:el_TF_trk}
}\\
\subfigure[]{
\includegraphics[width=0.45\textwidth]{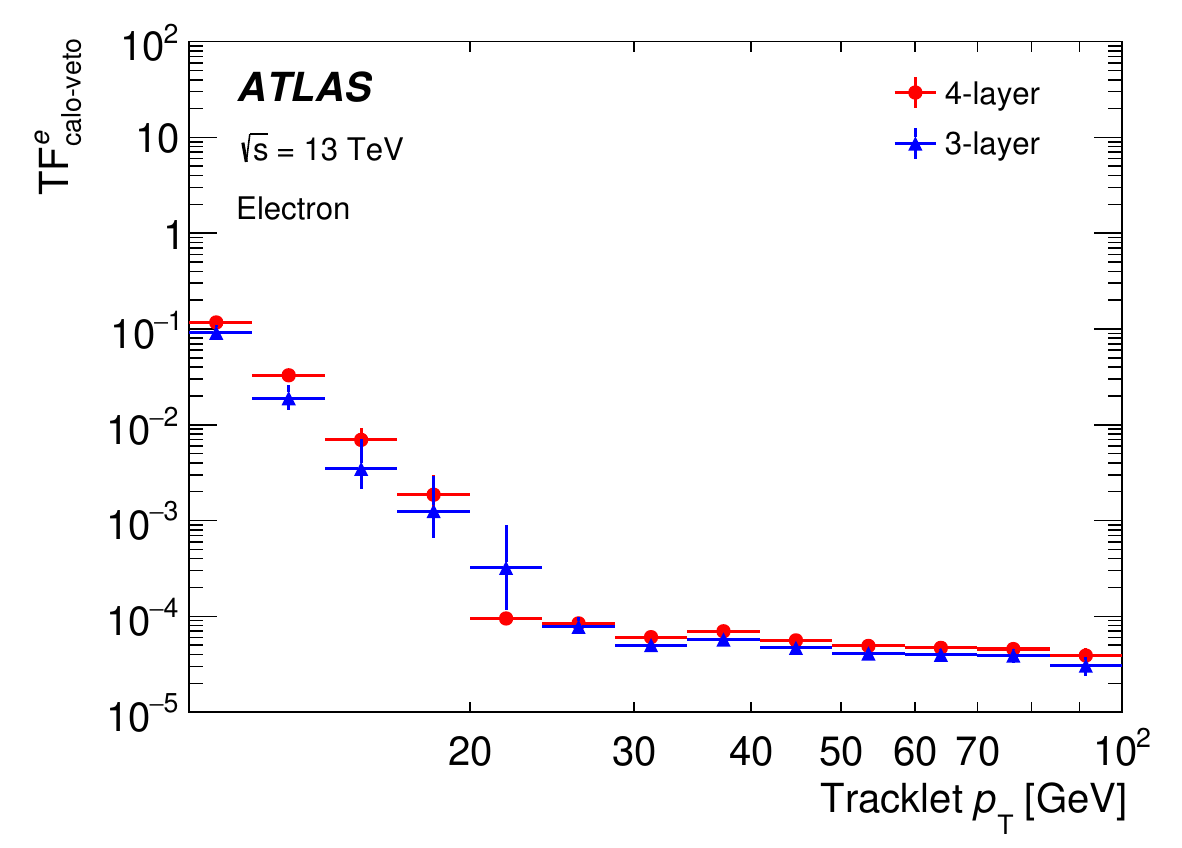}
\label{fig:bkg:el_TF_calo}
}
\subfigure[]{
\includegraphics[width=0.45\textwidth]{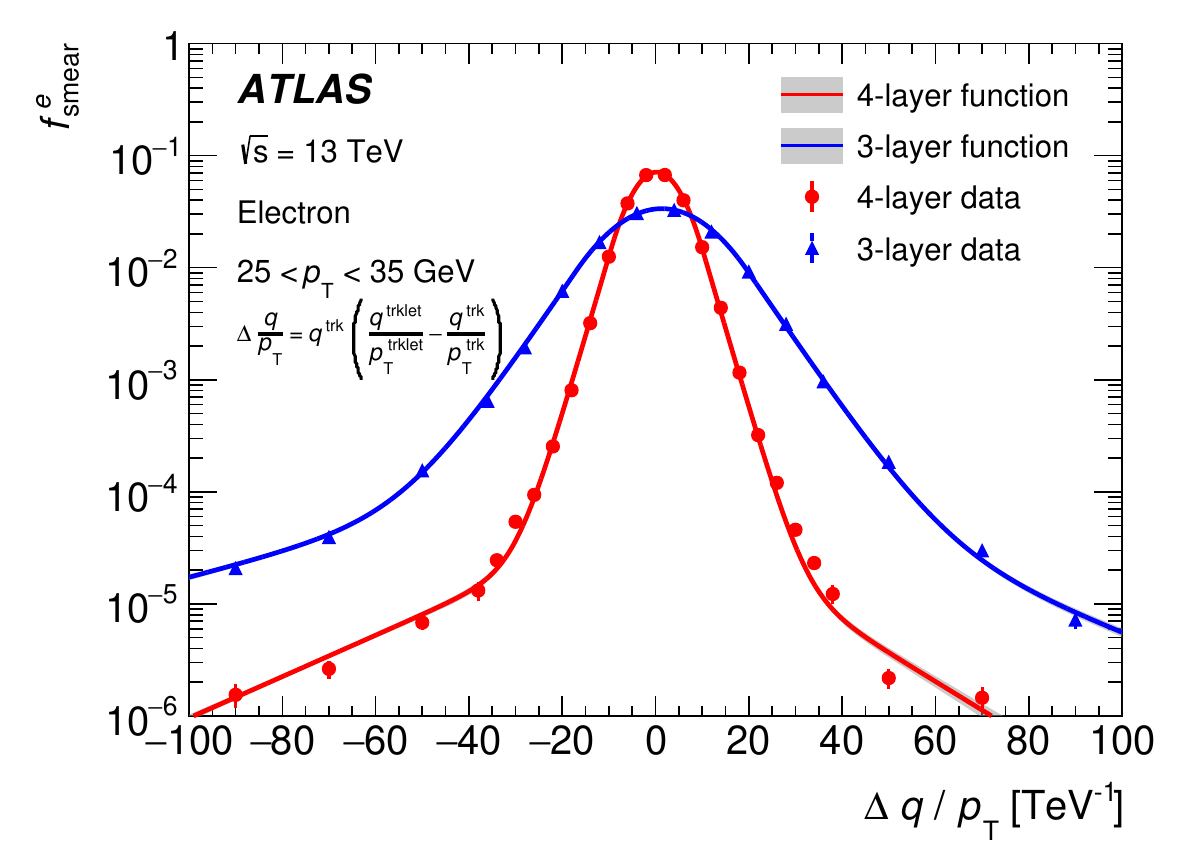}
\label{fig:bkg:el_smear}
}
\end{center}
\vspace{-0.75cm}
\caption{Individual contributions used in the estimation of the electron background: \protect\subref{fig:bkg:el_Template} presents the original electron \pt template, $N^{\mathrm{TRe}}(\pt)$, as a solid line, with the templates also shown after the application of each transfer/smearing factor as dashed lines; \protect\subref{fig:bkg:el_TF_trk} shows the transfer factor, $\mathrm{TF}^e_{\mathrm{tracklet}}$, used to correct for the likelihood of reconstructing an electron track as a tracklet; \protect\subref{fig:bkg:el_TF_calo} shows the transfer factors, $\mathrm{TF}^{e}_{\textrm{calo-veto}}$, used to correct for the likelihood of an electron not being associated with a calorimeter energy cluster; and \protect\subref{fig:bkg:el_smear} shows the smearing functions, $f^{e}_{\mathrm{smear}}$, used to correct the measured electron track \pt to the tracklet \pt (defined on the figure). While the tracklet \pt template and the TFs are dependent upon both \pt and $\eta$, the components are shown with their dependence upon the tracklet \pt distribution as a representative distribution. }
\end{figure}

which uses the yields in TRs modified by TFs and a smearing function that is described further below. Four electron-enriched TRs (TR$e$), with the same selections as the associated SRs but with the presence of exactly one electron instead of a tracklet, are used to extract a set of templates for the tracklet \pt spectrum ($N^{\mathrm{TR}e}(\pt,\eta)$). An example of the template \pt spectrum is presented in Figure~\ref{fig:bkg:el_Template} for the TR$e$ associated with \SRAh. The TR$e$ events must fire the single-electron trigger, with an offline $\pt$ requirement of $27\,\GeV$. To mimic the expected \met distribution in the SR, the electrons are removed from the \met calculation in this region, allowing the same \met selection as in the SR to be applied. A data-driven tag-and-probe method using $Z\rightarrow ee$ events is used to derive correction factors to estimate the likelihood that a standard electron is reconstructed as a tracklet ($\mathrm{TF}^e_{\mathrm{tracklet}}$) by identifying probe electrons that have a calorimeter energy cluster but no associated standard track. Figure~\ref{fig:bkg:el_TF_trk} shows the \pt dependence of the $\mathrm{TF}^e_{\mathrm{tracklet}}$ distributions. A further transfer factor ($\mathrm{TF}^{e}_{\textrm{calo-veto}}$) is calculated, again using a data-driven tag-and-probe method in $Z\rightarrow ee$ events, to estimate the likelihood that an electron with a standard track does not have an associated calorimeter energy cluster. Figure~\ref{fig:bkg:el_TF_calo} shows the \pt dependence of $\mathrm{TF}^{e}_{\textrm{calo-veto}}$ for three- and four-layer tracklets, where both TFs are calculated in bins of tracklet \pt and $\eta$. A transfer factor ($\mathrm{TF}^{e}_{\mathrm{tracklet+\pi}}$) is calculated and used specifically for the regions where the electron is associated with a low-energy charged-pion track and is used instead of $\mathrm{TF}^e_{\mathrm{tracklet}}$ when considering the regions requiring the presence of a pion. The application of these TFs allows the electron \pt template taken from the TR to be scaled in a way that takes into account the likelihood that an electron may not satisfy both of the tracking and calorimeter requirements, and thus is reconstructed as a tracklet. To account for the differences between a standard reconstructed electron track and a reconstructed tracklet, a $p_{\mathrm{T}}$-dependent smearing function ($f^{e}_{\mathrm{smear}}$) is applied. To calculate the smearing function, the tracklet reconstruction is applied to a standard track associated with an electron (again using $Z\rightarrow ee$ data) and the difference between the charge and \pt values of the standard track and tracklet are calculated. This function is fitted using a Gaussian core with exponential tails. Figure~\ref{fig:bkg:el_smear} shows the smearing functions used for three- and four-layer tracklets.

A similar method is used to constrain the muon backgrounds. The tracklet \pt distribution due to muons in a SR is calculated as

\begin{equation*}
f_{\mathrm{SR}}^{\mu}(\pt,\eta) = N^{\mathrm{TR}\mu}(\pt,\eta) \times \mathrm{TF}^\mu_{\mathrm{tracklet}}(\pt,\eta) \times \mathrm{TF}^\mu_{\textrm{MS-veto}}(\pt,\eta)\times f^{\mu}_{\mathrm{smear}}(\pt).
\end{equation*}

This is used in the same way as for the electron background and the individual terms are derived in a similar manner as discussed below. Four muon-enriched TRs (TR$\mu$) are defined and the tracklet \pt template is taken from events passing the same selection as in the SRs but containing exactly one muon instead of a tracklet. The TR$\mu$ events must fire the single-muon trigger, with an offline $\pt$ requirement of $27\,\GeV$. As with the electrons, the expected \met distribution in the SR is emulated by removing the muons from the \met calculation. Correction factors to account for muons reconstructed as tracklets ($\mathrm{TF}^\mu_{\mathrm{tracklet}}$), and muons that do not have an associated track in the MS ($\mathrm{TF}^{\mu}_{\textrm{MS-veto}}$), are calculated in bins of tracklet \pt and $\eta$ using a tag-and-probe method in $Z\rightarrow \mu\mu$ events. A transfer factor ($\mathrm{TF}^{\mu}_{\mathrm{tracklet+\pi}}$) is calculated specifically for the cases where the muon is associated with a low-energy charged-pion track and is used instead of $\mathrm{TF}^\mu_{\mathrm{tracklet}}$ in the regions requiring the presence of a pion. A $p_{\mathrm{T}}$-dependent smearing function ($f^{\mu}_{\mathrm{smear}}$) is applied to account for differences between the measured muon track $p_{\mathrm{T}}$ in the TR and the reconstructed tracklet \pt.

The final background considered in the analysis, the one due to hadronic interactions, is calculated as

\begin{equation*}
f_{\mathrm{SR}}^{\mathrm{had}}(\pt,\eta) = N^{\mathrm{TRhad}}(\pt,\eta) \times \mathrm{TF}^{\mathrm{had}}_{\mathrm{tracklet}}(\pt,\eta) \times \mathrm{TF}^{\mathrm{had}}_{\textrm{calo-veto}}(\pt,\eta) \times f^{\mathrm{had}}_{\mathrm{smear}}(\pt).
\end{equation*}

Similarly to the other backgrounds, yields in TRs are modified by TFs and a smearing function, which is described below. Four hadronic TRs (TRhad) are defined to extract the tracklet \pt template ($N^{\mathrm{TRhad}}(\pt,\eta)$) by applying the same selections as the associated SR, but with changed tracklet reconstruction requirements. In the TRs the track with the highest $\pt$ is required to have at least 14 hits in the TRT detector, at least five hits in the SCT, and be matched, within $\Delta R < 0.2$, to a calorimeter energy cluster with $E_{\mathrm{T}}^{\mathrm{clus}} > 3\,\GeV$, and $E_{\mathrm{T}}^{\mathrm{topoclus40}}/\pt^{\mathrm{track}} > 0.5$, resulting in a pure sample of hadronic tracks. Due to the unavailability of a data-driven tag-and-probe method to calculate the transfer factors, pion and electron \enquote{particle gun} simulation is used to estimate the probability to misidentify a hadron as a short track ($\mathrm{TF}^\mathrm{had}_{\mathrm{tracklet}}$) by calculating the ratio of hadron MC events reconstructed as tracklets to electron MC events reconstructed as tracklets and multiplying this by the $\mathrm{TF}^e_{\mathrm{tracklet}}$ value calculated from data as described above. A similar procedure is followed to calculate $\mathrm{TF}^\mathrm{had}_{\mathrm{tracklet+\pi}}$, which is subsequently used instead of $\mathrm{TF}^\mathrm{had}_{\mathrm{tracklet}}$ to calculate the SR contribution containing a low-energy pion. A transfer factor $\mathrm{TF}_{\textrm{calo-veto}}^{\mathrm{had}}$ is applied to estimate the likelihood of reconstructing a hadron without an associated calorimeter energy cluster; it is calculated as the ratio of MC events containing a tracklet associated with a calorimeter energy cluster within $\Delta R=0.4$ of the tracklet (when requiring $E_{\mathrm{T}}^{\mathrm{clus}} > 10\,\GeV$) to those without (when requiring $E_{\mathrm{T}}^{\mathrm{clus}}$ < 5\,GeV). Both TFs are dependent upon \pt and $\eta$. A $p_{\mathrm{T}}$-dependent smearing function ($f^{\mathrm{had}}_{\mathrm{smear}}$) is applied to take into account differences between the hadronic track and the reconstructed tracklet.

\begin{table}[tp]
\centering
\caption{Summary of the event and tracklet selection requirements for the CRs associated with the four-layer SRs. \label{tab:CRA_Selections}}
\scriptsize
\begin{tabular}{l | c | c | c | c }
\toprule
Variable & \CRAl & \CRAh & \CRAls & \CRAms \\
\midrule
\met trigger & \multicolumn{4}{c}{Passed} \\
Number of leptons &   \multicolumn{4}{c}{$= 0$} \\
Number of jets ($\pt > 20\,\GeV$) &   \multicolumn{4}{c}{$\geq 1$} \\
Leading jet $\pt$ &   \multicolumn{4}{c}{$> 100\,\GeV$} \\
$|\Delta\phi(j_1,\vec{p}_{\mathrm{T}}^{\mathrm{miss}})|$ &   \multicolumn{4}{c}{$> 1.0$} \\
Number of four-layer tracklets &   \multicolumn{4}{c}{$\geq 1$} \\
Tracklet $\pt$ &  $> 20\,\GeV$ &  $20{-}60\,\GeV$ & $> 20\,\GeV$ & $20{-}60\,\GeV$ \\
Tracklet $E_\mathrm{T}^{\mathrm{clus}}$ &  $< 5\,\GeV$ &  $< 5\,\GeV$ & $> 5\,\GeV$ & $> 5\,\GeV$ \\
\met & $< 150\,\GeV$ & $>300\,\GeV$ & $<150\,\GeV$ & $150{-}300\,\GeV$ \\
\bottomrule
\end{tabular}
\end{table}

After evaluating the tracklet \pt distributions for the electron, muon, hadron and fake backgrounds, they are used as the input to a binned likelihood fit (as described in Section \ref{sec:result}) to provide a final estimate of the background distributions in the SRs, performing an extrapolation to higher values of \met. The hadronic background tracklet distributions are normalised using \enquote{sideband} CRs: \CRAls and \CRAms (defined in Table~\ref{tab:CRA_Selections}) for the four-layer SRs, and \CRBols and \CRBoms (defined in Table \ref{tab:CR1p_Selections}) for both three-layer SRs, where the \enquote{S} in the superscript denotes the sideband CR for the region. Applying this normalisation results in a negligible dependence on MC when calculating the hadronic transfer factors. Figure~\ref{fig:background:FitSchematic} presents a schematic overview of the general background estimation strategy with the usage of CRs to normalise the hadronic background while the remaining backgrounds are taken from their templates as described above.

\begin{table}[b]
\centering
\caption{Summary of the event and tracklet selection requirements for the CRs associated with the three-layer SRs that contain a low-energy charged pion. \label{tab:CR1p_Selections}}
\scriptsize
\begin{tabular}{l | c | c | c | c | c }
\toprule
Variable & \CRBol & \CRBom & \CRBoh & \CRBols & \CRBoms \\
\midrule
\met trigger & \multicolumn{5}{c}{Passed} \\
Number of leptons &   \multicolumn{5}{c}{$= 0$} \\
Number of jets ($\pt > 20\,\GeV$) &   \multicolumn{5}{c}{$\geq 1$} \\
Leading jet $\pt$ &   \multicolumn{5}{c}{$> 100\,\GeV$} \\
$|\Delta\phi(j_1,\vec{p}_{\mathrm{T}}^{\mathrm{miss}})|$ &   \multicolumn{5}{c}{$> 1.0$} \\
Number of three-layer tracklets &   \multicolumn{5}{c}{$\geq 1$} \\
Number of charged pions & \multicolumn{5}{c}{$\geq 1$} \\
Tracklet $\pt$ &  $> 20\,\GeV$ &  $20{-}60\,\GeV$ & $20{-}60\,\GeV$ & $> 20\,\GeV$ & $20{-}60\,\GeV$ \\
Tracklet $E_\mathrm{T}^{\mathrm{clus}}$ &  $< 5\,\GeV$ &  $< 5\,\GeV$ & $< 5\,\GeV$ & $> 5\,\GeV$ & $> 5\,\GeV$ \\
\met & $< 150\,\GeV$ & $150{-}240\,\GeV$ & $>240\,\GeV$ & $< 150\,\GeV$ & $150{-}240\,\GeV$ \\
\bottomrule
\end{tabular}
\end{table}

\begin{table}[tp]
\centering
\caption{Summary of the event and tracklet selection requirements for the CRs associated with the three-layer SRs that veto a low-energy charged pion. \label{tab:CR1z_Selections}}
\scriptsize
\begin{tabular}{l | c | c | c  }
\toprule
Variable & \CRBzl & \CRBzm & \CRBzh \\
\midrule
\met trigger & \multicolumn{3}{c}{Passed} \\
Number of leptons &   \multicolumn{3}{c}{$= 0$} \\
Number of jets ($\pt > 20\,\GeV$) &   \multicolumn{3}{c}{$\geq 1$} \\
Leading jet $\pt$ &   \multicolumn{3}{c}{$> 100\,\GeV$} \\
$|\Delta\phi(j_1,\vec{p}_{\mathrm{T}}^{\mathrm{miss}})|$ &   \multicolumn{3}{c}{$> 1.0$} \\
Number of three-layer tracklets &   \multicolumn{3}{c}{$\geq 1$} \\
Number of charged pions & \multicolumn{3}{c}{$= 0$} \\
Tracklet $\pt$ &  $> 20\,\GeV$ &  $20{-}60\,\GeV$ & $20{-}60\,\GeV$\\
Tracklet $E_\mathrm{T}^{\mathrm{clus}}$ &  $< 5\,\GeV$ &  $< 5\,\GeV$ & $< 5\,\GeV$\\
\met & $< 150\,\GeV$ & $150{-}280\,\GeV$ & $>280\,\GeV$ \\
\bottomrule
\end{tabular}
\end{table}

\begin{figure}[t!]
\centering
\includegraphics[width=0.85\textwidth]{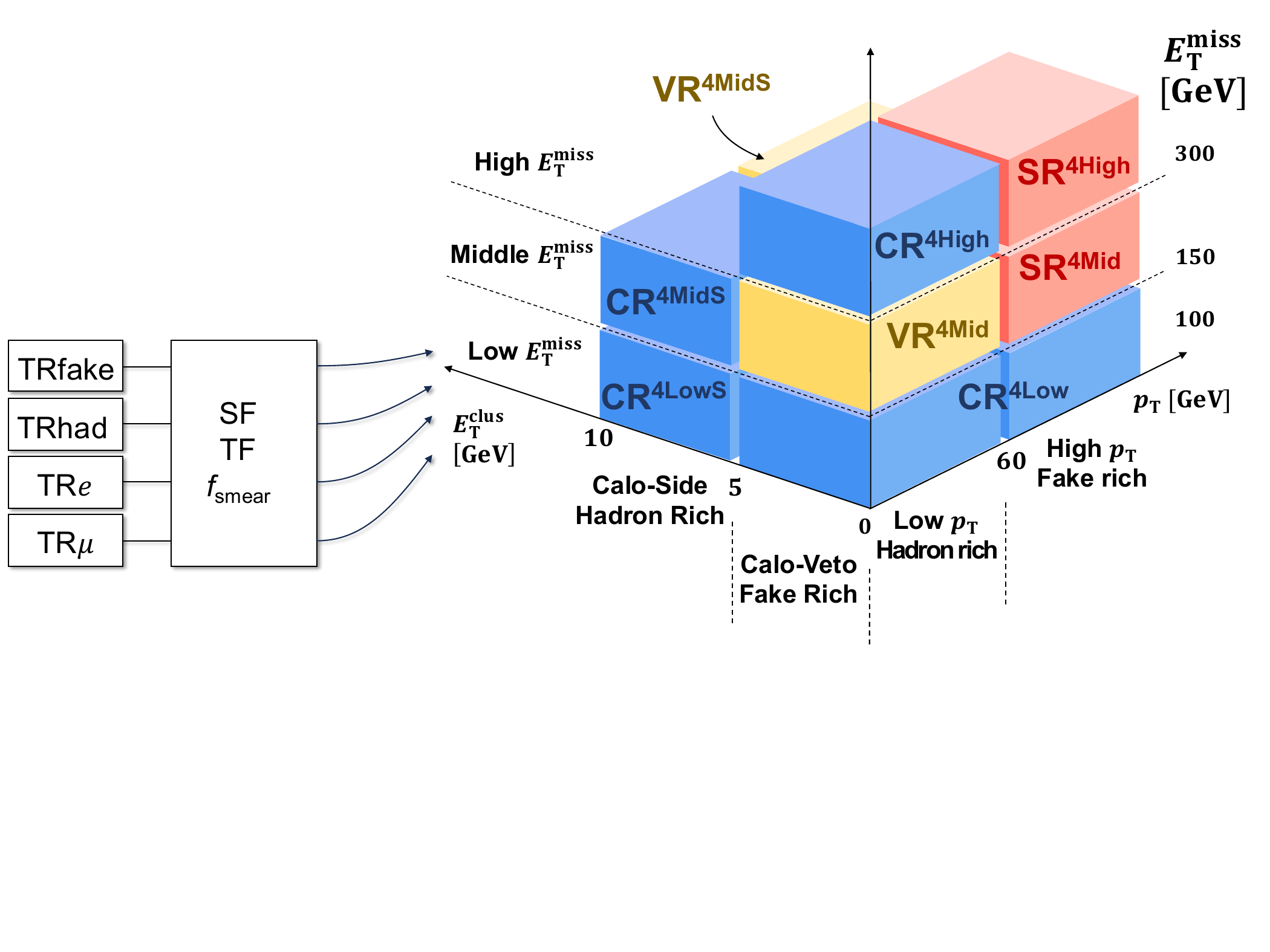}
\vspace{-3.5cm}
\caption{Schematic diagram for the background estimation strategy, describing the usage of template regions (TRs), and relaxed selections to calculate transfer factors (TFs) before performing the fit using control regions (CRs). The kinematic regions are orthogonal due to selections on \met, tracklet $p_{\mathrm{T}}$, and tracklet $E_\mathrm{T}^{\mathrm{clus}}$, with high calorimeter energies (\enquote{calo-side}) dominated by the hadron background, and low calorimeter energies (\enquote{calo-veto}) dominated by the fake background.}
\label{fig:background:FitSchematic}
\end{figure}

To evaluate the fit procedure and background modelling, orthogonal validation regions (VRs) are constructed by selecting events kinematically closer to the SR, and an extrapolation of the CR backgrounds to the VRs is performed. Five VRs are defined, as shown in Table~\ref{tab:VR_Selections}, with reversed selections on the tracklet $\pt$, tracklet $E_\mathrm{T}^{\mathrm{clus}}$, or $\met$ in order to remain orthogonal to both the SRs and CRs.

\begin{table}[tp]
\centering
\caption{Summary of the event and tracklet selection requirements for the validation regions. \label{tab:VR_Selections}}
\scriptsize
\begin{tabular}{l | c | c | c | c | c}
\toprule
Variable & \VRAm & \VRAms & \VRBom & \VRBoms & \VRBzm \\
\midrule
\met trigger & \multicolumn{5}{c}{Passed} \\
Number of leptons &   \multicolumn{5}{c}{$= 0$} \\
Number of jets ($\pt > 20\,\GeV$) &   \multicolumn{5}{c}{$\geq 1$} \\
Leading jet $\pt$ &   \multicolumn{5}{c}{$> 100\,\GeV$} \\
$|\Delta\phi(j_1,\vec{p}_{\mathrm{T}}^{\mathrm{miss}})|$ &   \multicolumn{5}{c}{$> 1.0$} \\
Number of four-layer tracklets &   \multicolumn{2}{c|}{$\geq 1$} & \multicolumn{3}{c}{= 0} \\
Number of three-layer tracklets &   \multicolumn{2}{c|}{-} & \multicolumn{3}{c}{$\geq 1$} \\
Number of charged pions & \multicolumn{2}{c|}{-} & \multicolumn{2}{c|}{$\geq 1$} & $= 0$ \\
Tracklet $\pt$ & $< 60\,\GeV$ & $> 60\,\GeV$ & $> 60\,\GeV$ & $> 60\,\GeV$ & $> 60\,\GeV$ \\
Tracklet $E_\mathrm{T}^{\mathrm{clus}}$ &  $< 5\,\GeV$ &  $> 5\,\GeV$ & $< 5\,\GeV$ & $> 5\,\GeV$ & $< 5\,\GeV$ \\
\met & $150{-}300\,\GeV$ & $150{-}300\,\GeV$ & $150{-}240\,\GeV$ & $150{-}240\,\GeV$ & $150{-}280\,\GeV$ \\
\bottomrule
\end{tabular}
\end{table}

\FloatBarrier

% End of text imported from the .//sections/backgrounds.tex input file

%
\section{Systematic uncertainties}
\label{sec:Systs}
%

% The next lines are included from the .//sections/systematics.tex input file
Systematic uncertainties in the background estimates are driven by the uncertainties in the transfer factors, normalisations and shape of the tracklet \pt templates for the electron, muon, hadron and fake backgrounds. These uncertainties are dominated by the statistical uncertainties of the data in regions used to derive the background estimates, and are always below 10\%. When calculating values of parameters which are based on data, such as the TFs and smearing functions, the statistical uncertainty of the data is used as a $\pm 1\sigma$ variation. When performing the statistical fit, the values of the transfer factors and the individual bins of each tracklet \pt template are allowed to vary within their statistical uncertainties. Each source of background uncertainty (\pt template statistical uncertainties, TFs, and smearing functions) is treated as correlated across each set of associated CRs and SRs but uncorrelated between backgrounds, except $\mathrm{TF}^\mathrm{had}_{\mathrm{tracklet}}$, which is correlated with $\mathrm{TF}^e_{\mathrm{tracklet}}$ since both TFs are used in the estimation of the hadronic background. No dedicated uncertainty is associated with the overall background estimation methodology as, within other sources of uncertainties including the statistics of the data, the predicted post-fit tracklet \pt distributions in the VRs are consistent with the data.

Several sources of experimental uncertainty in the SUSY signal estimates are considered. The uncertainties are included in the likelihood fit as Gaussian nuisance parameters that affect the expectation values for each SR bin. Jet energy scale and resolution uncertainties are derived as a function of the jet $\pt$, $\eta$, and flavour, using data and simulated events, as detailed in Ref.~\cite{JETM-2018-05}.
Lepton reconstruction and identification uncertainties~\cite{EGAM-2021-01,MUON-2018-03} are included in the fit, but are found to have negligible impact. Uncertainties in the tracklet reconstruction efficiency due to dead or inefficient modules are estimated using a tag-and-probe method comparing $Z\rightarrow \mu\mu$ MC and data. An $\eta$-dependent difference of up to 10\% between data and MC is found and this inefficiency is added as an uncertainty. A 6\% uncertainty is applied due to selection differences between data and MC $V$+\;\!jets events. An uncertainty related to the low-energy charged pions is also applied, combining uncertainties of 10\% in the reconstruction efficiency, taken from Ref.~\cite{IDTR-2021-03}, and 6\% in the BDT selection efficiency, evaluated by varying the BDT input parameters by 1$\sigma$ and examining the effect on the output score. All uncertainties in the final-state object reconstruction are propagated to the $\met$ calculation, including an additional term taking into account uncertainties in the scale and resolution of the soft term~\cite{JETM-2020-03}. In the \SRAm region, the \met selection used is in the \enquote{turn-on} region of the \met trigger where the trigger is found to be between 85\% and 90\% efficient for the signal event topology. Therefore, a trigger efficiency uncertainty is also applied, as an uncertainty in the signal yield.

The theoretical uncertainties originating from the choice of factorisation and renormalisation scales for the SUSY signal samples are evaluated by varying the corresponding MC generator parameters by a factor of two around their nominal values. The uncertainties associated with the choice of PDF set~\cite{Ball:2014uwa} and the value of the strong coupling constant are also considered. Finally, five variations of the A14 tune of \PYTHIA are used to account for ISR modelling effects due to the underlying event, jet structure, and different aspects of extra jet production. The total uncertainty from experimental sources is found to be ${\sim}5$\% for all signal scenarios. The theoretical modelling uncertainties contribute up to 20\%, increasing with increasing $\chinoonepm$/$\tau$-slepton mass, and are primarily driven by the ISR modelling uncertainty.

Figure~\ref{fig:systematicsUnc} presents a breakdown of systematic uncertainties affecting the background estimates in the SRs, grouped by the source of the uncertainty. The total uncertainty may not simply be the sum in quadrature of the individual uncertainties, because the fit can induce correlations between the components. In all SRs, the dominant uncertainty comes from the normalisation of the fake background, and the subdominant uncertainty in most regions arises from the shape of the fake background. Due to the limited number of tracklets available for the tracklet \pt templates in the \SRBo region, the statistical uncertainty of the template shape contributes significantly to the background uncertainty in this region.

\begin{figure}[bt]
\centering
\includegraphics[width=0.55\textwidth]{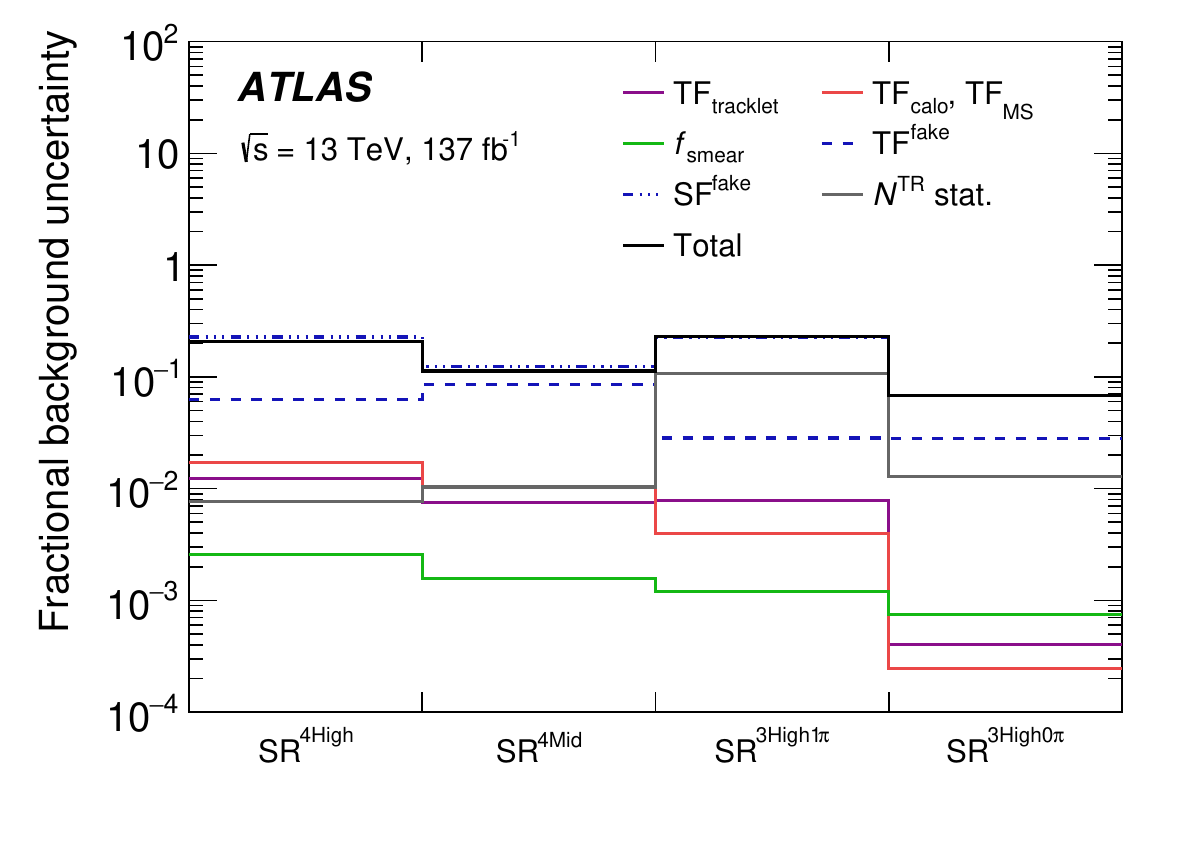}
\vspace{-0.75cm}
\caption{A summary of the systematic uncertainties affecting the background estimates in the SRs after the likelihood fit to data in the CRs is performed (\enquote{Post-fit}). The total uncertainty is indicated by a solid black line. The individual uncertainties may not sum in quadrature to the total uncertainty, due to correlations between the different components. The uncertainties arising from the estimation of the fake-tracklet background, which are the dominant uncertainties, are shown as blue dashed lines.}
\label{fig:systematicsUnc}
\end{figure}

% End of text imported from the .//sections/systematics.tex input file

%
\section{Results}
\label{sec:result}
%

% The next lines are included from the .//sections/results.tex input file
Three profile likelihood fits are performed: to assess the accuracy of the SM background determination ({\em background-only} fit), to compute the $p$-value of the SM-only hypothesis ({\em model-independent} fit), and to evaluate the confidence level (CL) for excluding a specific new-physics hypothesis ({\em model-dependent} fit)~\cite{Baak:2014wma}. Due to the low event yields in the SRs, pseudoexperiments are used for statistical interpretation of the results when performing the model-independent fit, while the asymptotic approximation~\cite{Cowan:2010js} is used when performing the model-dependent fit.

When performing the background-only fit, the hadronic background yield is allowed to float in the CRs, resulting in a set of normalisation factors which are uncorrelated between SRs. One hadronic normalisation factor is calculated for each SR and is propagated as a general scaling factor to the SR's tracklet \pt distribution. Only the CRs, and not the SRs, are used to constrain the SM background. The electron, muon and fake-tracklet background yields are allowed to vary within their uncertainties when the fit is performed, but no overall normalisation is derived for these backgrounds.

Figure~\ref{fig:results:BinnedCRs} presents an overview of the CRs after the background-only fit. The CRs that are used to normalise the hadronic backgrounds have, by design, a ratio of data to background event yield of one, whereas the other CRs that are used to derive the tracklet \pt distribution for the fake background are not normalised and the data--MC difference is applied as the $\mathrm{SF}_{\met}^{\mathrm{fake}}$ value from the fake-background estimation procedure before performing the fit.

\begin{figure}[t]
\centering
\includegraphics[width=0.75\textwidth]{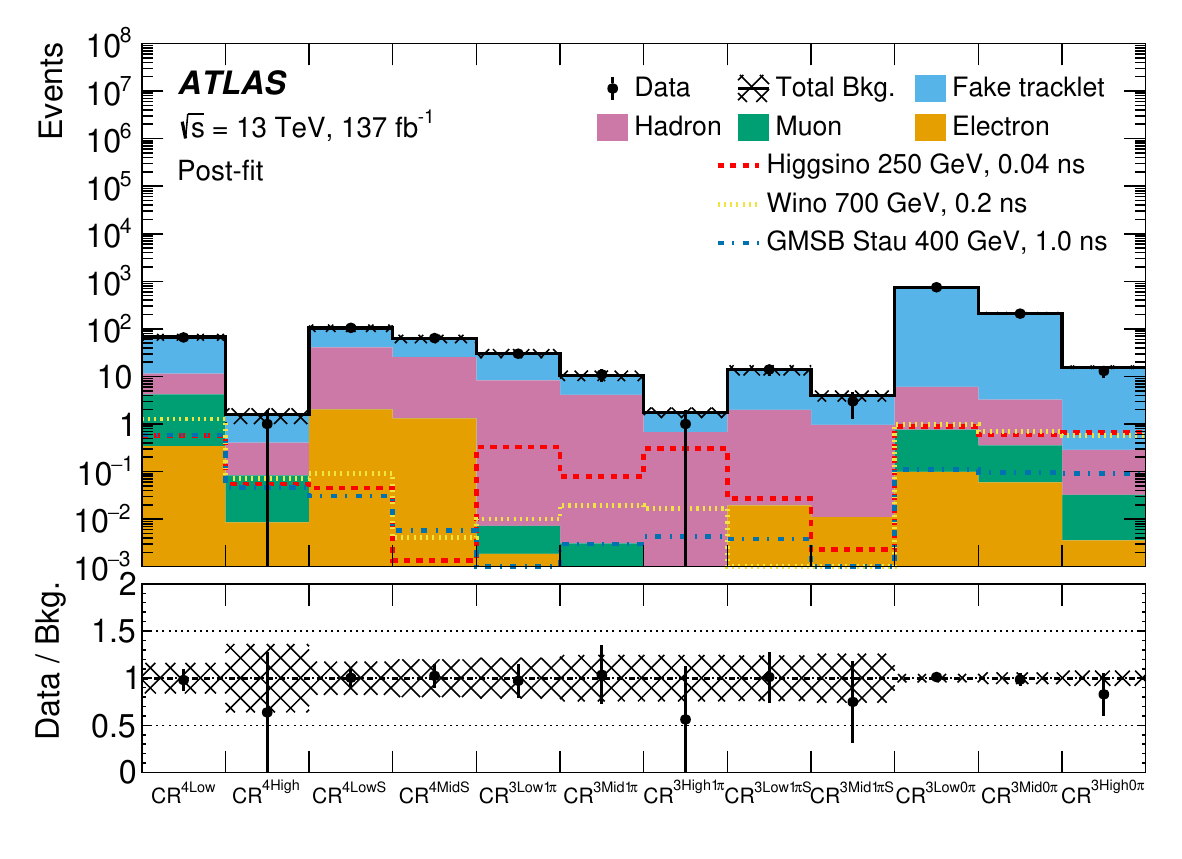}
\vspace{-0.5cm}
\caption{The post-fit yields in the CRs. All statistical and systematic uncertainties are considered in the error bands. Three representative pre-fit signal yields are indicated with dashed lines.}
\label{fig:results:BinnedCRs}
\end{figure}

To evaluate the fit procedure and background modelling, the CR normalisations are extrapolated to the VRs and the results are compared with data and validated before performing the same extrapolation to the SRs. Good agreement is seen between the data and prediction when examining the modelling of the tracklet $\pt$ shape in three of the VRs as shown in Figure~\ref{fig:VR_N1}. The tracklet \pt distribution differs between the electron and muon backgrounds because an electron is more likely to undergo bremsstrahlung than a muon.

\begin{figure}[htb]
\begin{center}
\subfigure[]{
\includegraphics[width=0.44\textwidth]{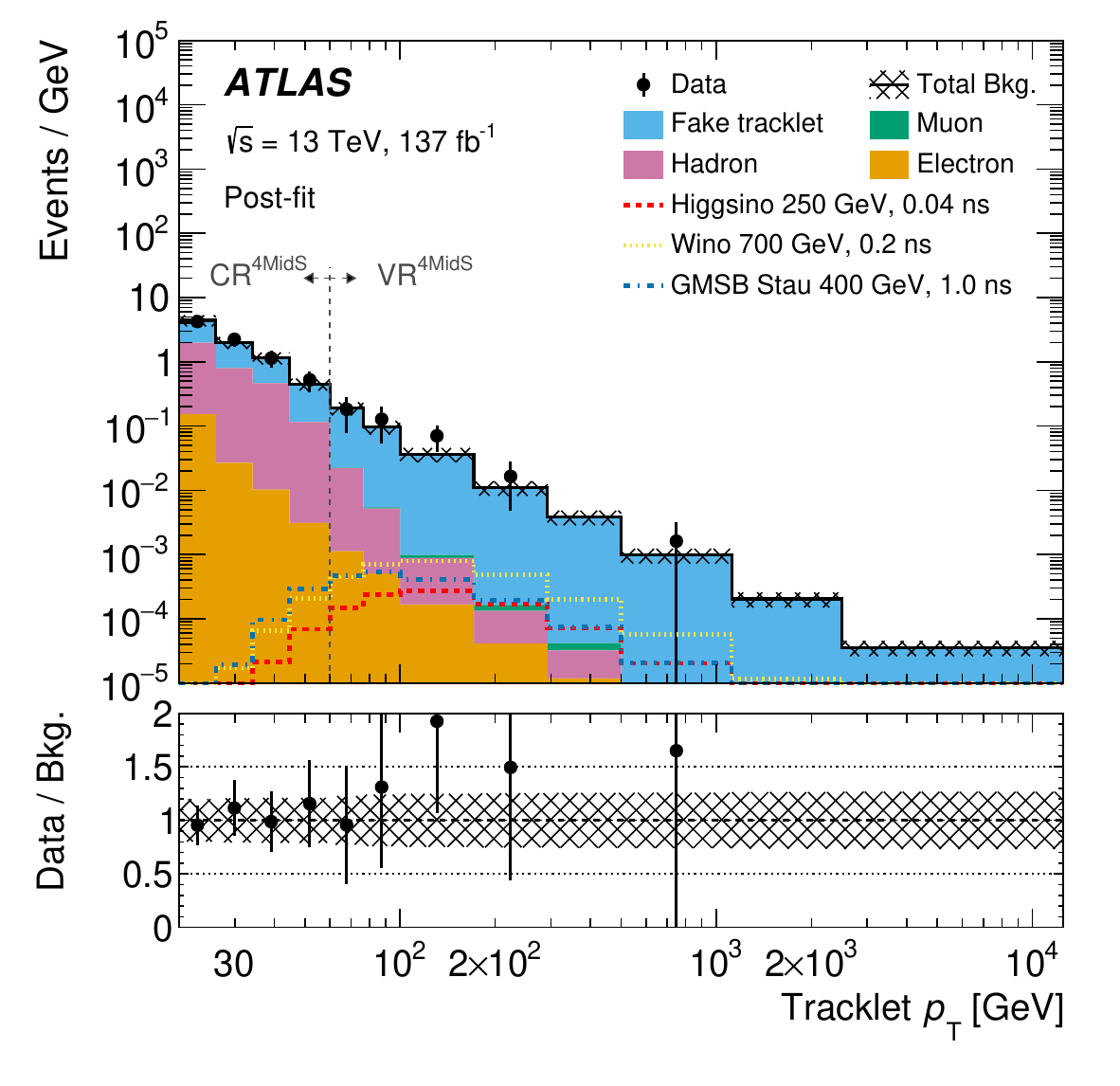}\label{fig:VR_N1:a}
}
\subfigure[]{
\includegraphics[width=0.44\textwidth]{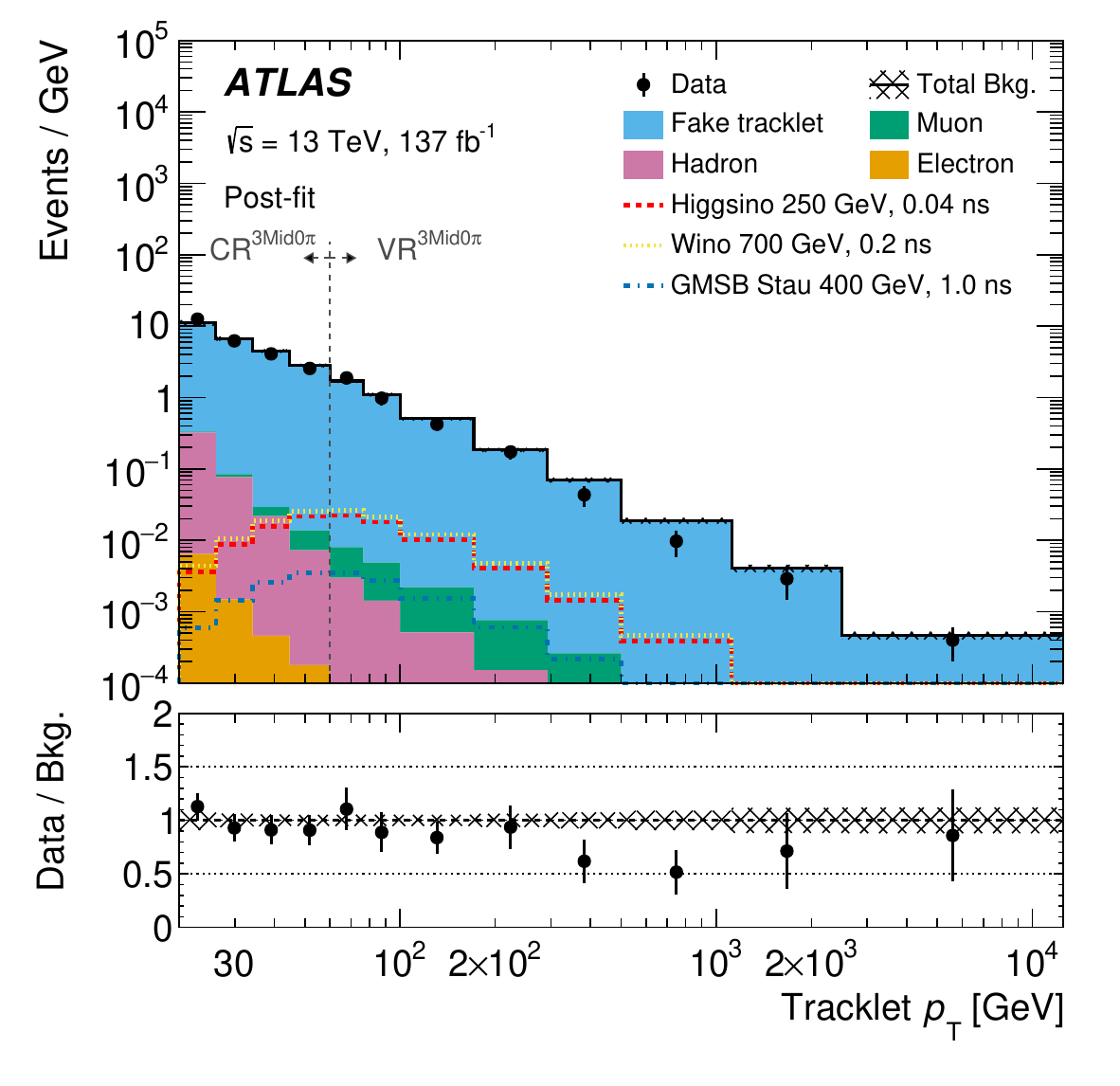}\label{fig:VR_N1:b}
}
\subfigure[]{
\includegraphics[width=0.44\textwidth]{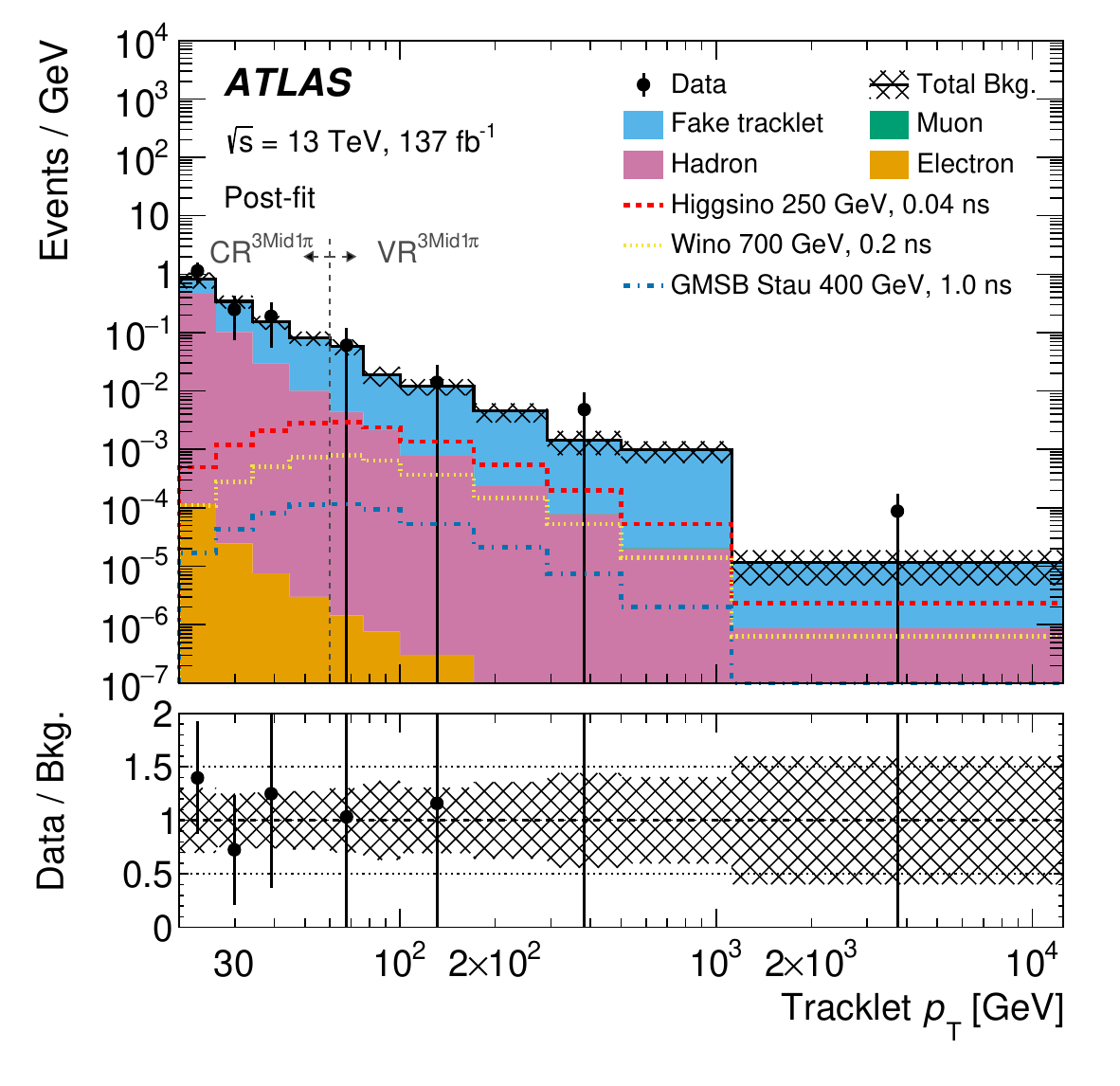}\label{fig:VR_N1:c}
}\\
\end{center}
\vspace{-0.75cm}
\caption{The post-fit tracklet \pt distributions in \protect\subref{fig:VR_N1:a} \VRAms, \protect\subref{fig:VR_N1:b} \VRBzm, and \protect\subref{fig:VR_N1:c} \VRBom. The dashed vertical line denotes the boundary between the CR and VR selections. The pre-fit signal distributions are indicated with dashed lines. The lower pad shows the ratio of the observed yields to the post-fit predicted background yields. The last bin includes overflow events. All statistical and systematic uncertainties are considered in the error bands.}
\label{fig:VR_N1}
\end{figure}

The level of post-fit agreement in the VRs and SRs is shown in Figure~\ref{fig:results:BinnedVRSR}, which also shows the statistical significance~\cite{Cousins:2007bmb} of the observed data yields and the background prediction in each of the regions. Good agreement is seen in the VRs. Small excesses are seen in \SRAh and \SRBo, corresponding to local significances of 1.9$\sigma$ and 1.0$\sigma$ respectively. These discrepancies and how the events compare with the previous analysis are discussed further at the end of this section.

\begin{figure}[t!]
\centering
\includegraphics[width=0.75\textwidth]{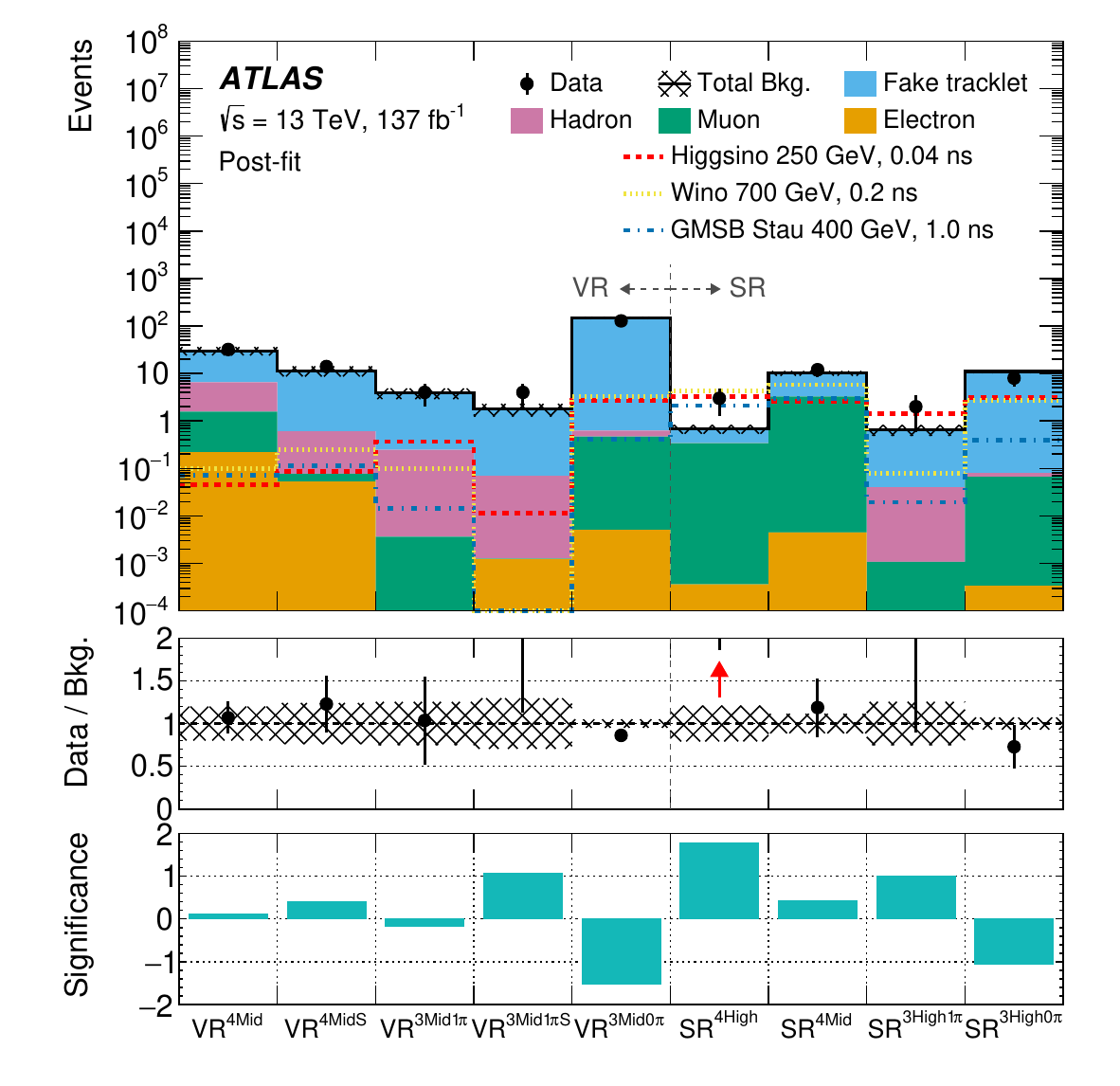}
\vspace{-0.5cm}
\caption{The observed and expected event yields in the VRs and SRs. The lower pads show the ratio of the observed yields to the post-fit predicted background yields and the statistical significance~\cite{Cousins:2007bmb} of the observed data. A red arrow is used to indicate that a ratio value is outside the displayed interval. The background estimates are calculated from the background-only fit. All statistical and systematic uncertainties are considered in the error bands. Three representative pre-fit signal yields are indicated with dashed lines.}
\label{fig:results:BinnedVRSR}
\end{figure}

Figure~\ref{fig:SR_N1} presents post-fit distributions of the tracklet \pt in each SR. In the regions where a relatively high number of background events are expected (\SRAm and \SRBz), the data are in good agreement with the background estimates, with the distributions agreeing within uncertainties. Due to the small number of events in \SRAh and \SRBo, less can be discerned from the level of agreement of the tracklet \pt distributions. Figure~\ref{fig:SRA:a} shows the presence of three data events in a single \SRAh bin, corresponding to tracklet \pt values of 139\,GeV, 142\,GeV and 152\,GeV.

\begin{figure}[b]
\begin{center}
\subfigure[]{
\includegraphics[width=0.44\textwidth]{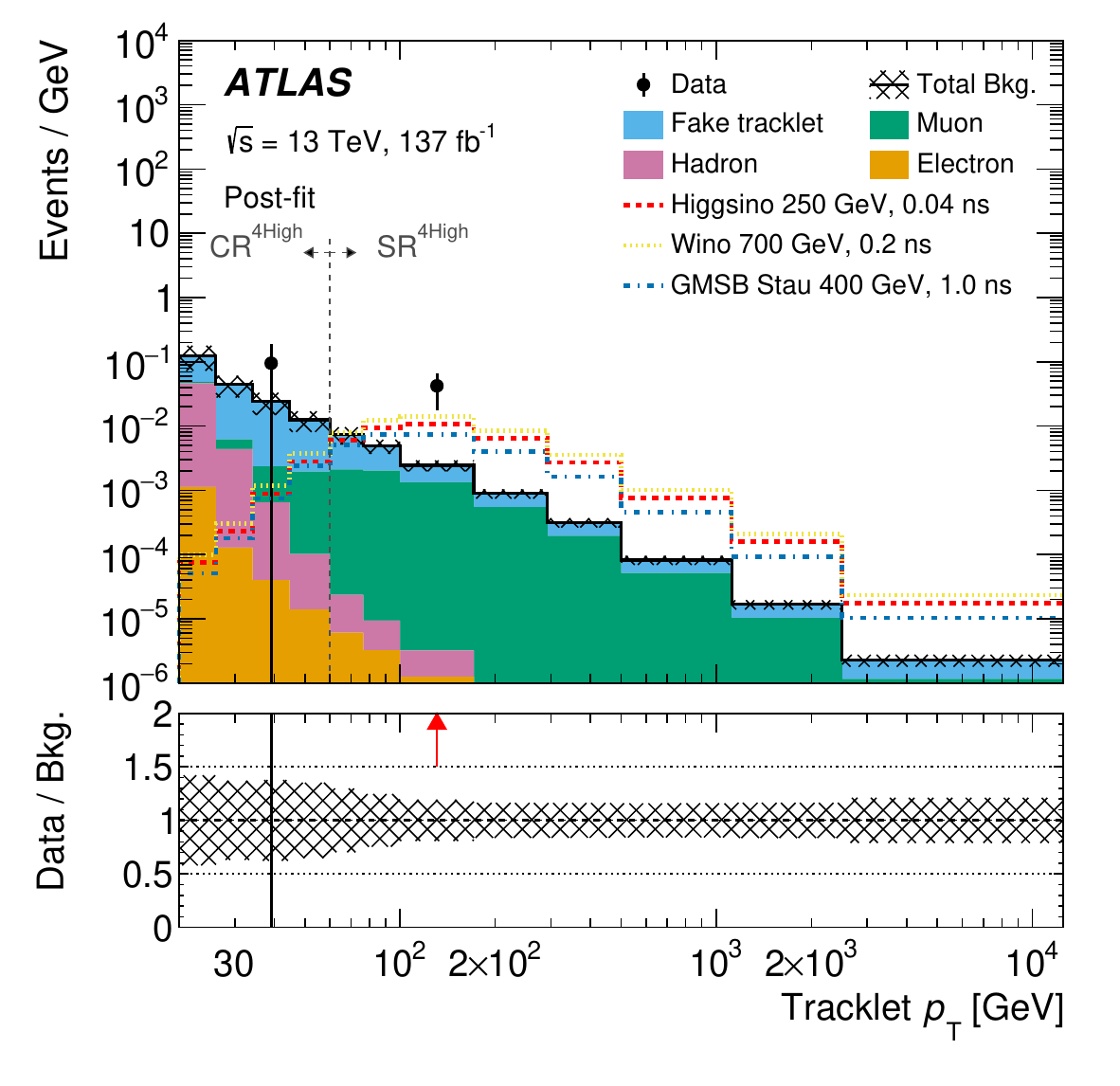}\label{fig:SRA:a}
}
\subfigure[]{
\includegraphics[width=0.44\textwidth]{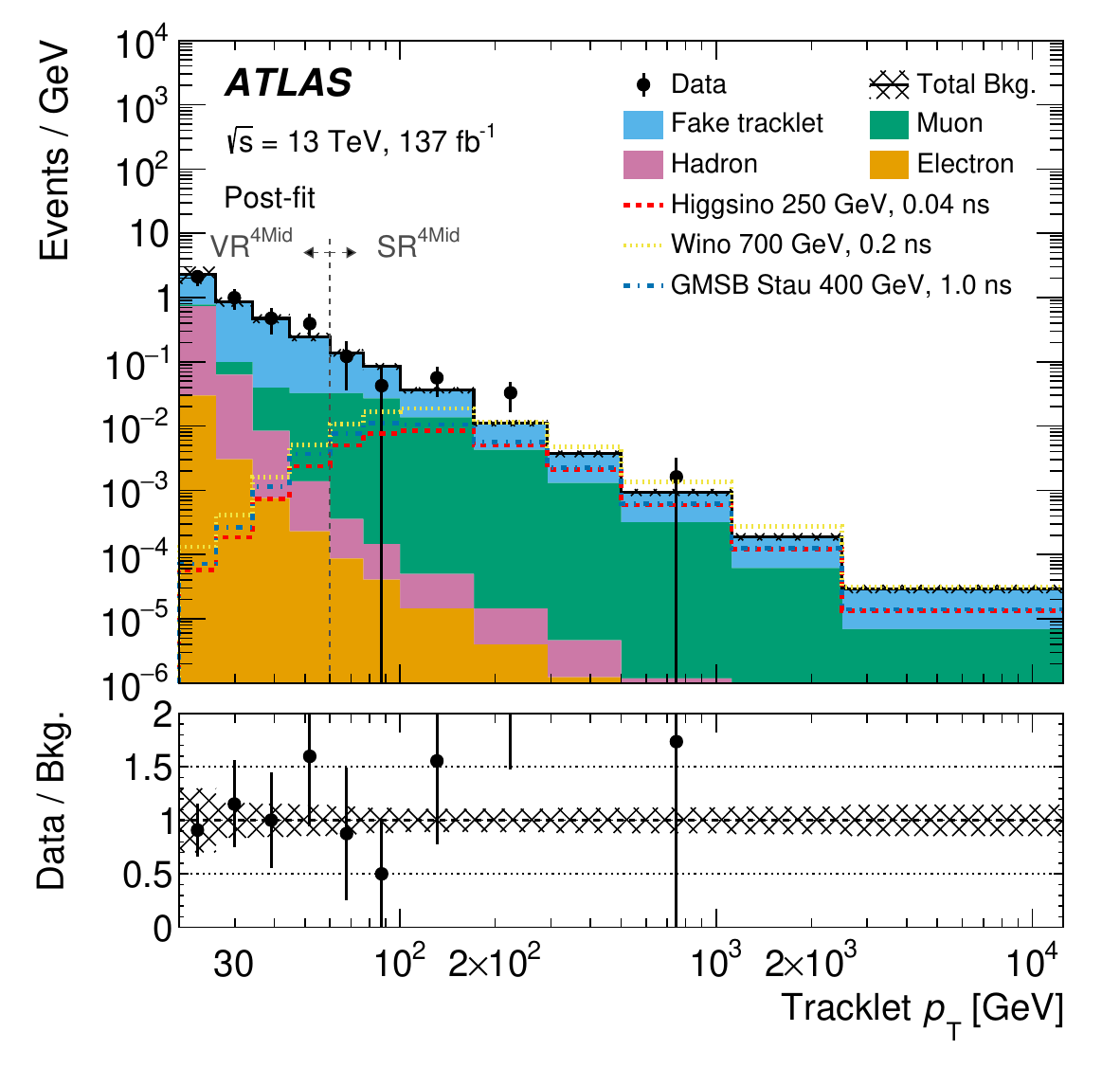}\label{fig:SRA:b}
}\\
\subfigure[]{
\includegraphics[width=0.44\textwidth]{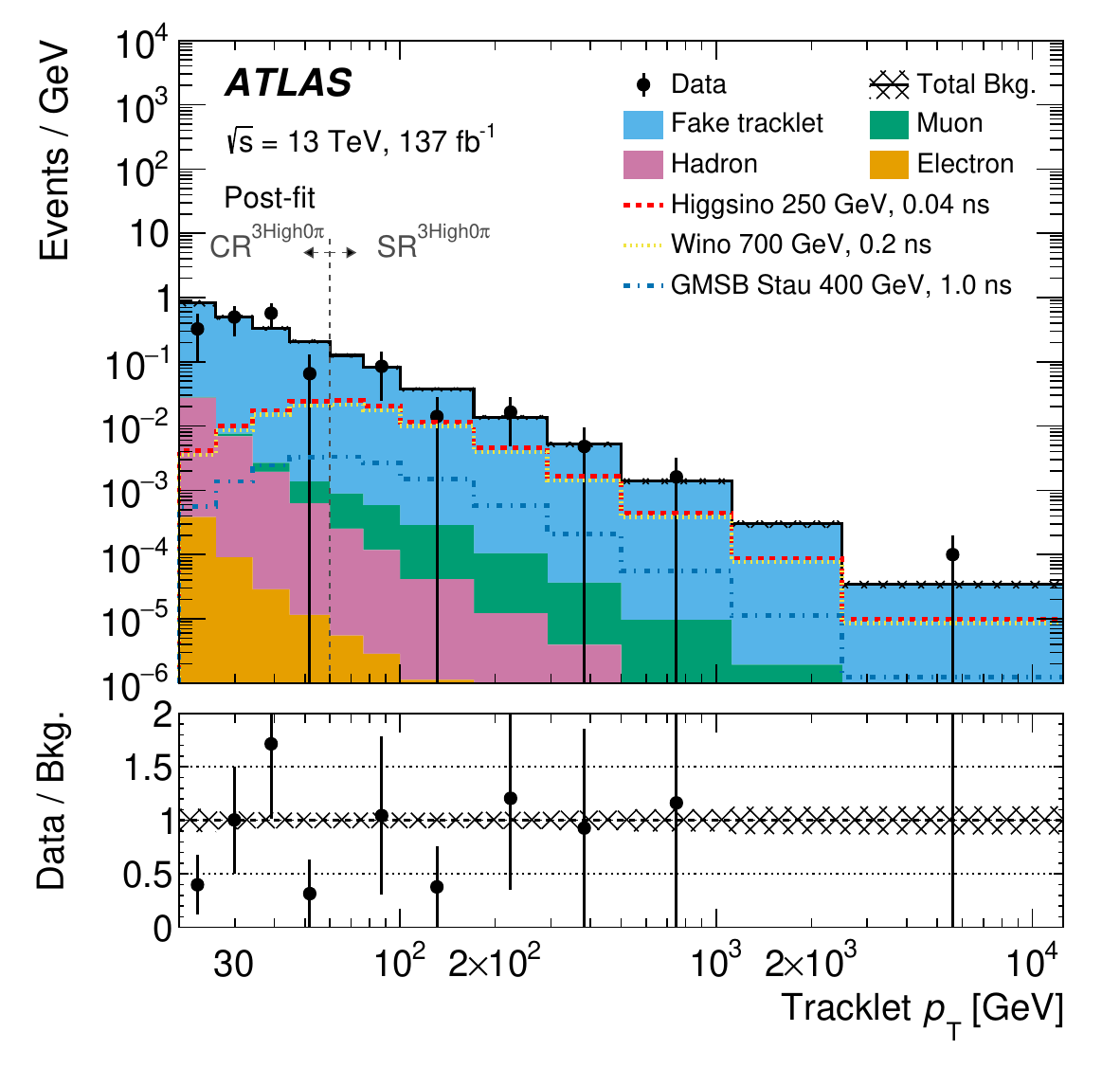}\label{fig:SRA:c}
}
\subfigure[]{
\includegraphics[width=0.44\textwidth]{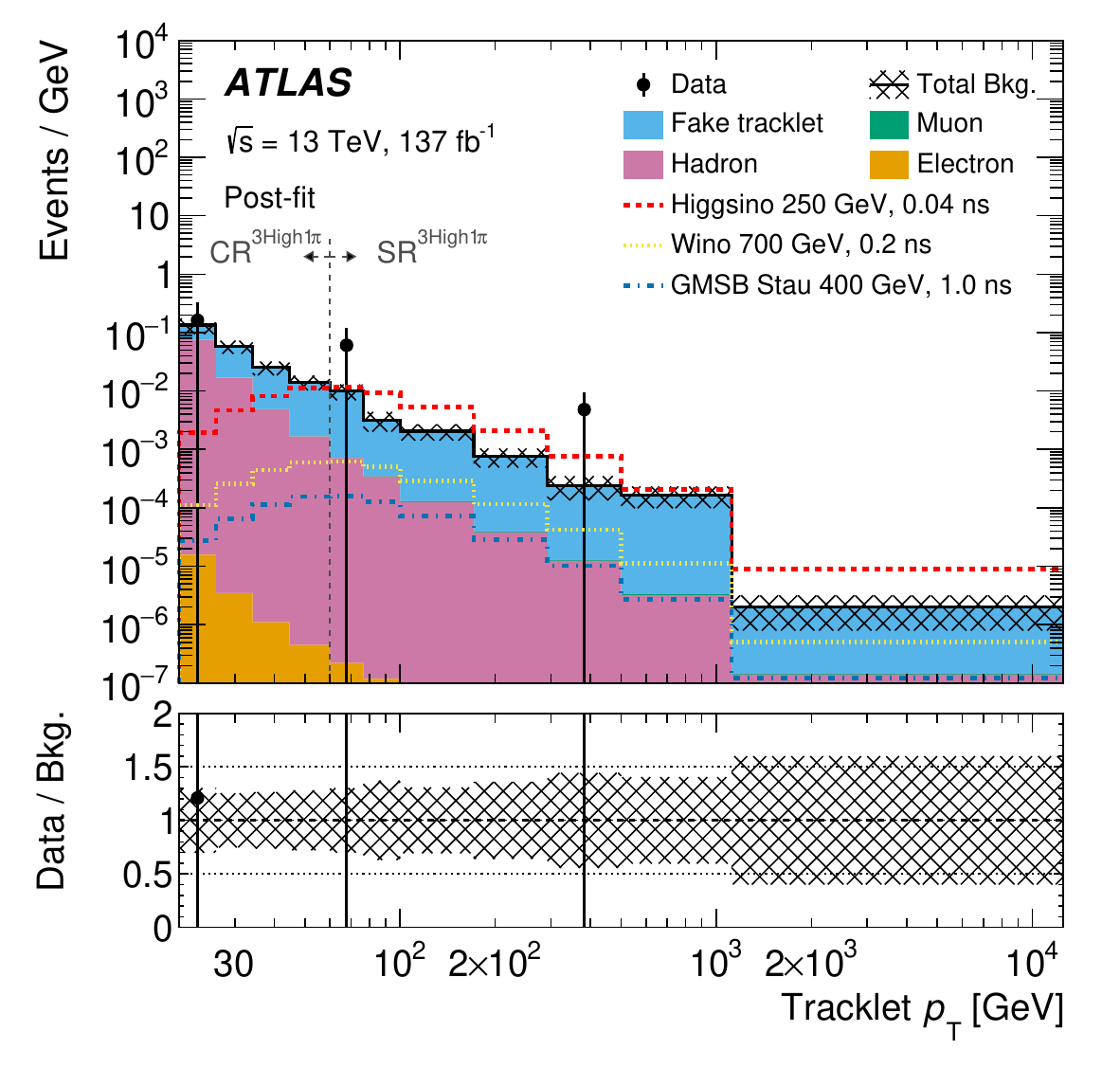}\label{fig:SRA:d}
}
\end{center}
\vspace{-0.5cm}
\caption{The post-fit tracklet \pt distributions in \protect\subref{fig:SRA:a} \SRAh, \protect\subref{fig:SRA:b} \SRAm, \protect\subref{fig:SRA:c} \SRBz, and \protect\subref{fig:SRA:d} \SRBo. The dashed vertical line denotes the boundary between the CR/VR and SR selections. The lower pad shows the ratio of the observed data yields to the post-fit predicted background estimates. A red arrow is used to indicate that a ratio value is outside the displayed interval. The last bin includes overflow events. All statistical and systematic uncertainties are considered in the error bands. Three representative pre-fit signal yields are indicated with dashed lines.
}
\label{fig:SR_N1}
\end{figure}

Table~\ref{tab:SRA_Yields} shows the observed and expected yields in the SRs, as well as the model-independent fit results in each SR. The model-independent fit uses the same CRs as the background-only fit. A single-bin SR is also included in the fit by summing the individual tracklet-\pt bins in a given region. The hadronic background is then allowed to float in the CRs and the single-bin SR. This method uses a profile-likelihood-ratio test statistic~\cite{Cowan:2010js}, where a generic new-physics signal of intensity $\mu_{\mathrm{sig}}$ is assumed to contribute exclusively to each individual, single-bin SR, and is used to assess the $p$-value of the background-only hypothesis (corresponding to $\mu_{\mathrm{sig}} = 0$) and to extract $S^{95}_{\mathrm{exp}}$ and $S^{95}_{\mathrm{obs}}$, i.e.\ the expected and observed 95\% CL limits on the number of signal events in the SR, using the CL$_\mathrm{s}$ prescription~\cite{Read:2002hq}. The $S^{95}_{\mathrm{obs}}$ limit is also converted to a limit on the visible cross-section $\sigma_{\mathrm{obs}}^{95}$ by dividing by the total integrated luminosity.

\begin{table}[hb]
\renewcommand\arraystretch{1.5}
\newcolumntype{M}{r@{${}\pm{}$}l}
\centering
\caption{Summary of the observed and post-fit expected SR yields, with \enquote{Bkg.} denoting background, and the model-independent fit results. All systematic and statistical uncertainties are considered. The calculated $p_0$-value is capped at 0.5 and shown together with the corresponding significance in standard deviations $(Z)$. \label{tab:SRA_Yields}}
\scriptsize
\begin{tabular}{l MMMM}
\toprule
& \multicolumn{2}{c}{\SRAh} & \multicolumn{2}{c}{\SRAm} & \multicolumn{2}{c}{\SRBo} & \multicolumn{2}{c}{\SRBz} \\
\midrule
Fake Bkg.     & $0.33$   & $0.14$   & $6.8$   & $1.1$   & $0.62$   & $0.16$   & $10.9$   & $0.8$ \\
Hadron Bkg.   & $(5$     & $4)\times 10^{-4}$ & $(7$ & $7)\times 10^{-3}$ & $0.029$ & $0.014$ & $0.011$ & $0.021$ \\
Electron Bkg. & $(3.7$   & $2.1)\times 10^{-4}$ & $(4.6$ & $2.7)\times 10^{-3}$ & $(10$ & $9)\times 10^{-6}$ & $(2.5$ & $1.3)\times 10^{-4}$ \\
Muon Bkg.     & $0.342$  & $0.015$  & $3.28$  & $0.14$  & $(1.1$   & $0.6)\times 10^{-3}$ & $0.068$ & $0.009$ \\
\midrule
Total Bkg.    & $0.68$   & $0.14$   & $10.1$  & $1.1$   & $0.65$   & $0.16$   & $11.0$   & $0.8$ \\
\midrule
Observed      & \multicolumn{2}{c}{$3$}  & \multicolumn{2}{c}{$12$} & \multicolumn{2}{c}{$2$} & \multicolumn{2}{c}{$8$} \\
\midrule
$p_{0}~(Z)$ & \multicolumn{2}{c}{0.033 (1.9)} & \multicolumn{2}{c}{0.26 (0.6)} & \multicolumn{2}{c}{0.14 (1.1)} & \multicolumn{2}{c}{0.5 (0)} \\
$S_{\mathrm{exp}}^{95}$ & \multicolumn{2}{c}{$3.7^{+0.8}_{-0.6}$} & \multicolumn{2}{c}{$12.3^{+3.6}_{-2.9}$} & \multicolumn{2}{c}{$3.2^{+1.0}_{-0.2}$} & \multicolumn{2}{c}{$8.3^{+2.8}_{-2.6}$} \\
$S_{\mathrm{obs}}^{95}$ & \multicolumn{2}{c}{6.0} & \multicolumn{2}{c}{12.4} & \multicolumn{2}{c}{5.0} & \multicolumn{2}{c}{5.8} \\
$\sigma_{\mathrm{obs}}^{95}~[\mathrm{fb}]$ & \multicolumn{2}{c}{0.044} & \multicolumn{2}{c}{0.090} & \multicolumn{2}{c}{0.037} & \multicolumn{2}{c}{0.043} \\
\bottomrule
\end{tabular}
\end{table}

In the absence of a significant excess in the observed data, 95\% CL exclusion limits are placed on the SUSY signal scenarios by statistically combining all SRs, binned in tracklet \pt, which provides greater sensitivity than simply considering only one SR at a time. The binning is chosen such that the low tracklet-\pt bins retain sufficient granularity to ensure fit stability and signal sensitivity, with wider bins used as tracklet \pt increases, reflecting the poorer tracklet \pt resolution and limited discrimination power of the highest bins. The reweighting procedure used to extrapolate signal scenarios to larger mass splittings could potentially affect the soft-pion kinematics. To prevent modelling issues related to changing soft-pion kinematics, the 3-layer SR selections are not applied to signals with lifetimes of greater than $0.2\,$ns. The range of mass splittings considered corresponds to a $200\,\MeV$ difference and this strict cut-off ensures that the soft-pion modelling is applied only to regions with low mass splittings and consistent pion kinematics. A model-dependent fit is used, where each signal scenario is introduced and a fit to the tracklet \pt distribution is performed, with CRs and SRs used to normalise the hadronic background and determine the signal strength for a specific signal model.
The four-layer regions dominate the sensitivity, and although the expected sensitivity in the SUSY models is significantly better than in the previous analysis~\cite{SUSY-2018-19}, by around $100\,\GeV$ in SUSY particle masses, the small excess in \SRAh weakens the observed exclusion limit.

The model-dependent limits for wino pair production are shown in Figure~\ref{fig:results:WinoLimits}, where wino-like charginos with lifetimes of $1\,$ns and masses up to $1020\,\GeV$ are expected to be excluded. The small excess observed in the signal regions results in a weaker observed mass exclusion limit of $880\,\GeV$.
Figure~\ref{fig:results:HiggsinoLimits} shows the model-dependent limits for higgsino pair production, where in the long-lifetime region of ${\sim}1$\,ns, the sensitivity extends up to masses of $720\,\GeV$ ($840\,\GeV$ expected).
Figures~\ref{fig:results:WinoLimits} and \ref{fig:results:HiggsinoLimits} also show the exclusion sensitivities when converting the chargino lifetime to the mass-splitting between the electroweakinos and when comparing it with the theoretical expectation for mass splitting in wino- or higgsino-like scenarios. In the wino-like case, charginos are excluded for masses up to $670\,\GeV$ ($770\,\GeV$ expected), while in the higgsino-like case, charginos are excluded for masses up to $225\,\GeV$ ($250\,\GeV$ expected), improving upon previous results by $15\,\GeV$. Compared to the previous analysis, this search is able to exclude smaller values of the signal cross-section. For example, when considering a higgsino signal scenario of $m(\chinoonepm) = 300$\,GeV and $\tau(\chinoonepm) = 0.05$\,ns the signal cross-section exclusion limit is a factor of 2.5 times lower than the previous result. When considering $m(\chinoonepm) = 500$\,GeV and $\tau(\chinoonepm) = 0.1$\,ns the signal cross-section exclusion limit is a factor of 1.8 lower than in the previous analysis.

\begin{figure}[tb!]
\begin{center}
\subfigure[]{
\includegraphics[width=0.45\textwidth]{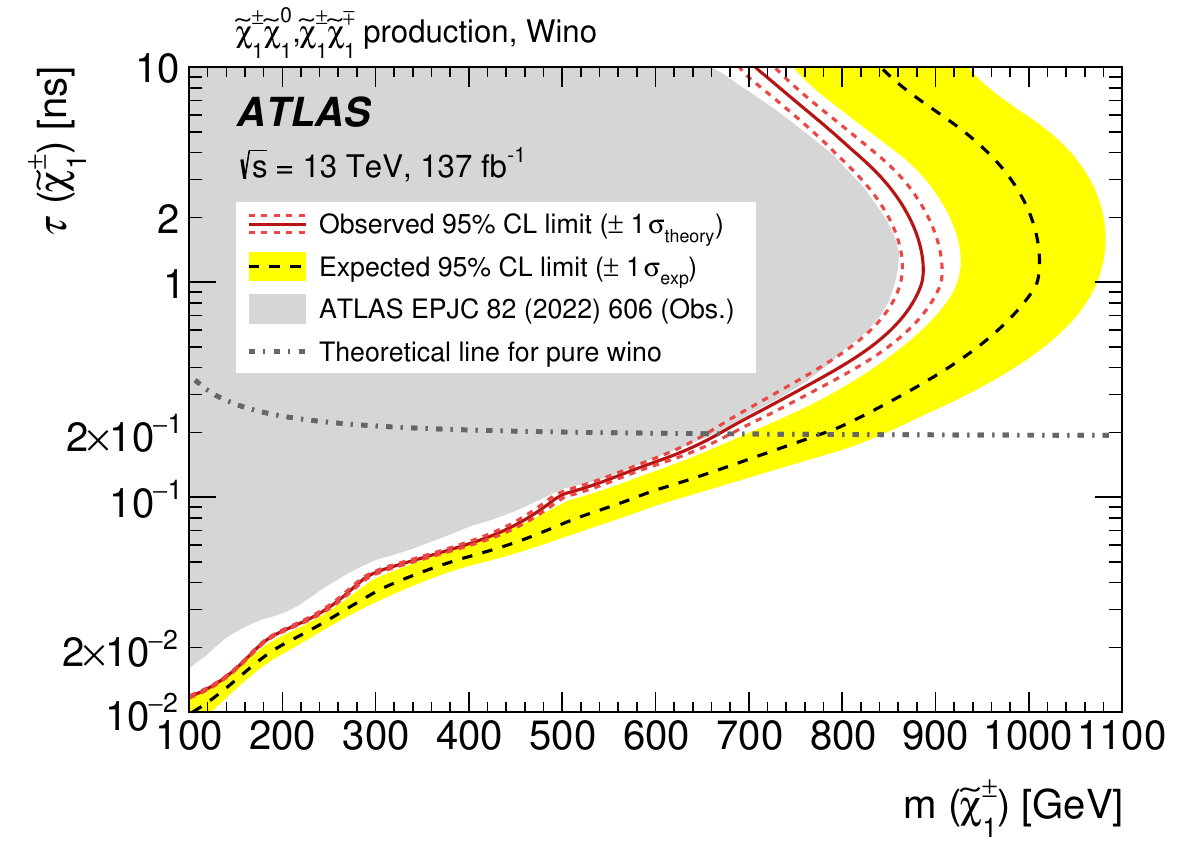}
\label{fig:results:WinoFull}
}
\subfigure[]{
\includegraphics[width=0.45\textwidth]{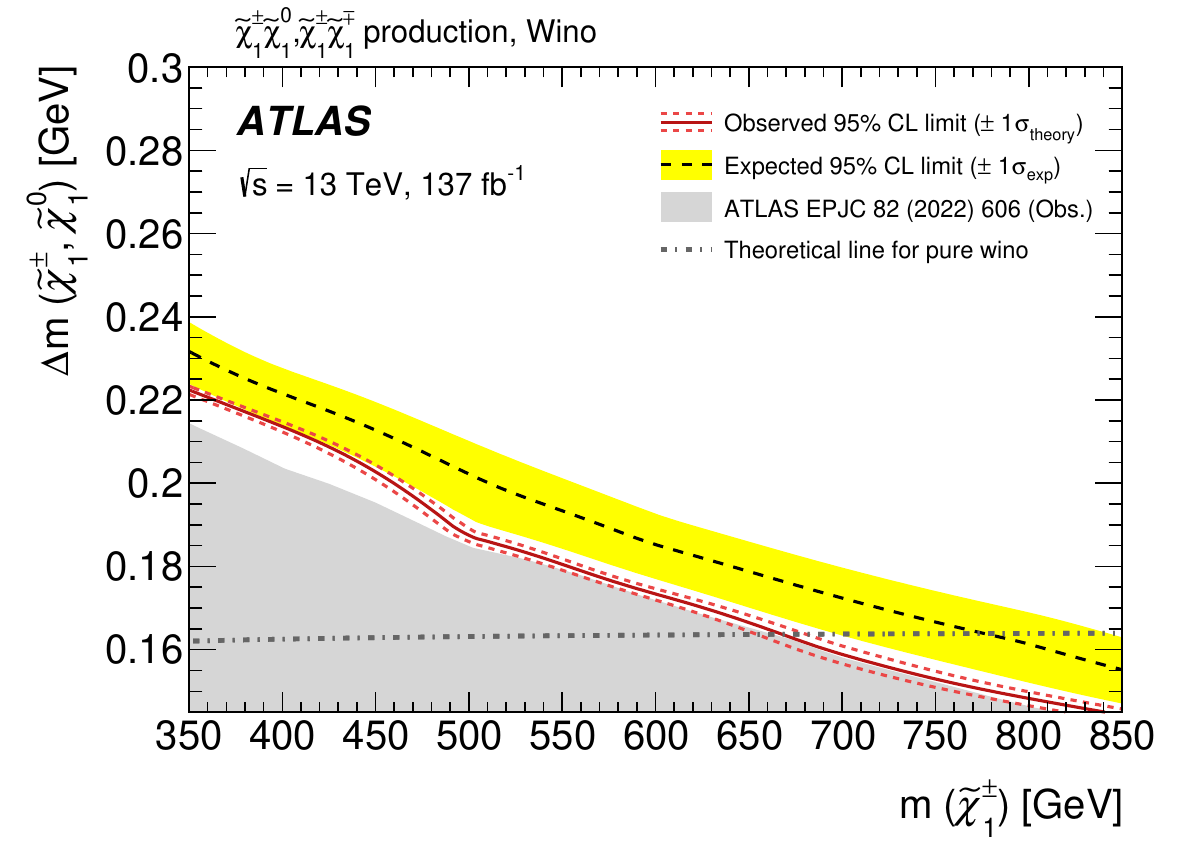}
\label{fig:results:WinoDM}
}
\end{center}
\vspace{-0.75cm}
\caption{The expected (dashed line) and observed (solid line) 95\% CL exclusion contours for wino production as a function of the chargino mass and \protect\subref{fig:results:WinoFull} lifetime, or \protect\subref{fig:results:WinoDM} mass splitting of the electroweakinos. The dotted lines indicate the $\pm \sigma_{\mathrm{theory}}$ from signal cross-section uncertainties, while the solid band indicates the $\pm \sigma_{\mathrm{exp}}$ from experimental systematic uncertainties and statistical uncertainties on the data yields. The solid grey area shows the region excluded by Ref.~\cite{SUSY-2018-19}. The dashed-dotted line is the prediction of the chargino lifetime or wino mass splitting as a function of the chargino mass as predicted by the SM loop corrections~\cite{Ibe:2012sx}. \label{fig:results:WinoLimits}}
\end{figure}

\begin{figure}[tb!]
\begin{center}
\subfigure[]{
\includegraphics[width=0.45\textwidth]{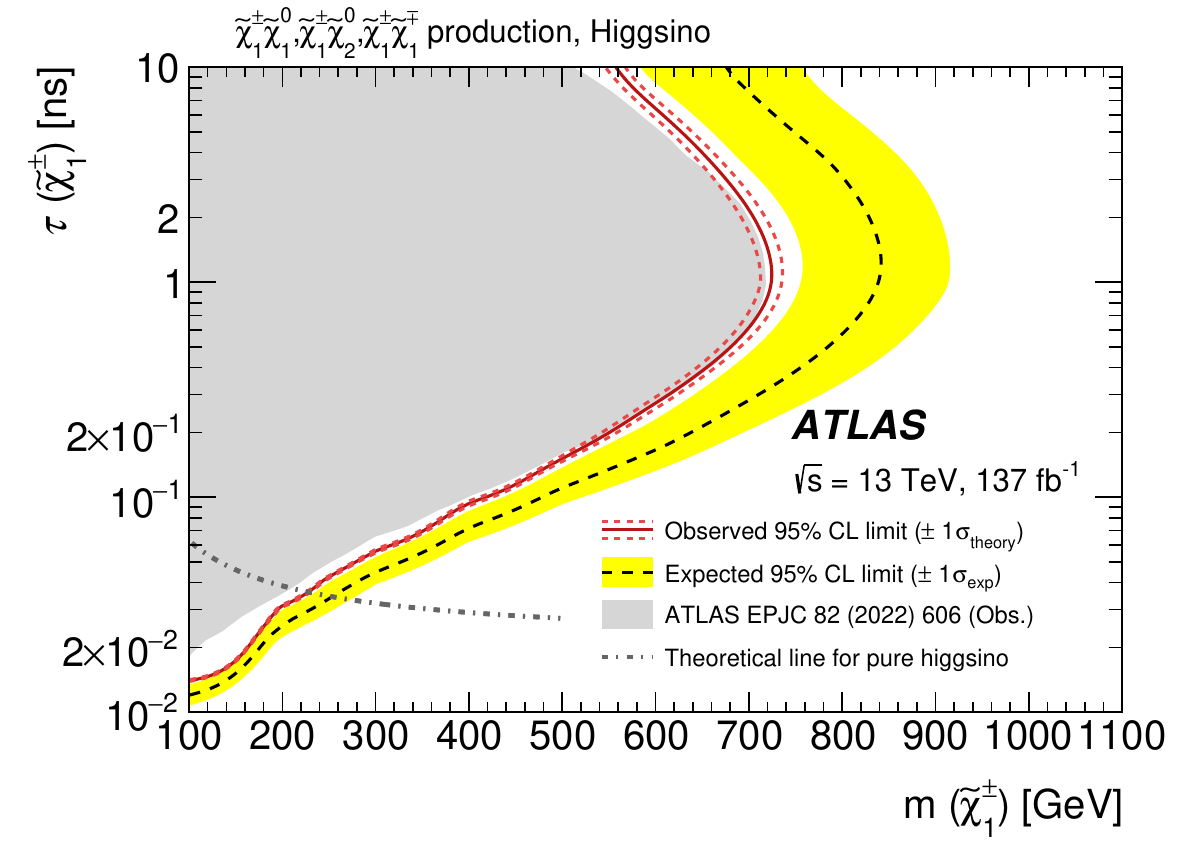}
\label{fig:results:HiggsinoFull}
}
\subfigure[]{
\includegraphics[width=0.45\textwidth]{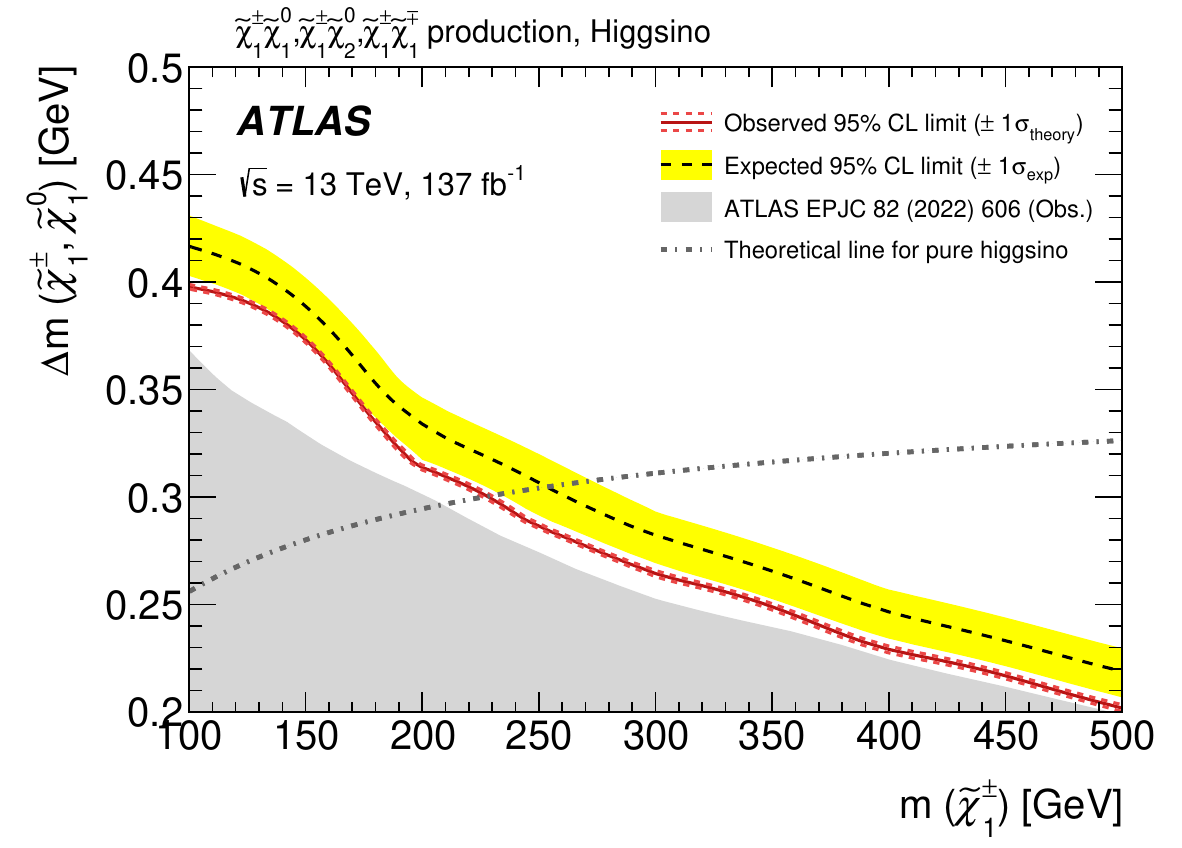}
\label{fig:results:HiggsinoDM}
}
\end{center}
\vspace{-0.75cm}
\caption{The expected (dashed line) and observed (solid line) 95\% CL exclusion contours for higgsino production as a function of the chargino mass and \protect\subref{fig:results:HiggsinoFull} lifetime, or \protect\subref{fig:results:HiggsinoDM} mass splitting of the electroweakinos. The dotted lines indicate the $\pm \sigma_{\mathrm{theory}}$ from signal cross-section uncertainties, while the solid band indicates the $\pm \sigma_{\mathrm{exp}}$ from experimental systematic uncertainties and statistical uncertainties on the data yields. The solid grey area shows the region excluded by Ref.~\cite{SUSY-2018-19}. The dashed-dotted line is the prediction of the chargino lifetime or higgsino mass splitting as a function of the chargino mass as predicted by the SM loop corrections~\cite{Thomas:1998wy}. \label{fig:results:HiggsinoLimits}}
\end{figure}

The 95\% CL exclusion limits in the $\tau$-slepton scenarios are presented in Figure~\ref{fig:results:StauLimits}. Due to the small cross-section for $\stau\stau$ production, the signal yield is too low in the three-layer SRs to offer sensitivity, so only the four-layer SRs and associated CRs are used in the model-dependent fit. The sensitivity in CMSSM- and GMSB-inspired scenarios is comparable, with $\tau$-sleptons excluded for masses up to around $320\,\GeV$ and $300\,\GeV$ ($390\,\GeV$ and $380\,\GeV$ expected), respectively, for a mass difference of ${\sim}2\,\GeV$ between the $\tau$-slepton and \ninoone/\gravino (corresponding to $\tau$-slepton lifetimes of ${\sim}1$\,ns).

\begin{figure}[tb!]
\begin{center}
\subfigure[]{
\includegraphics[width=0.45\textwidth]{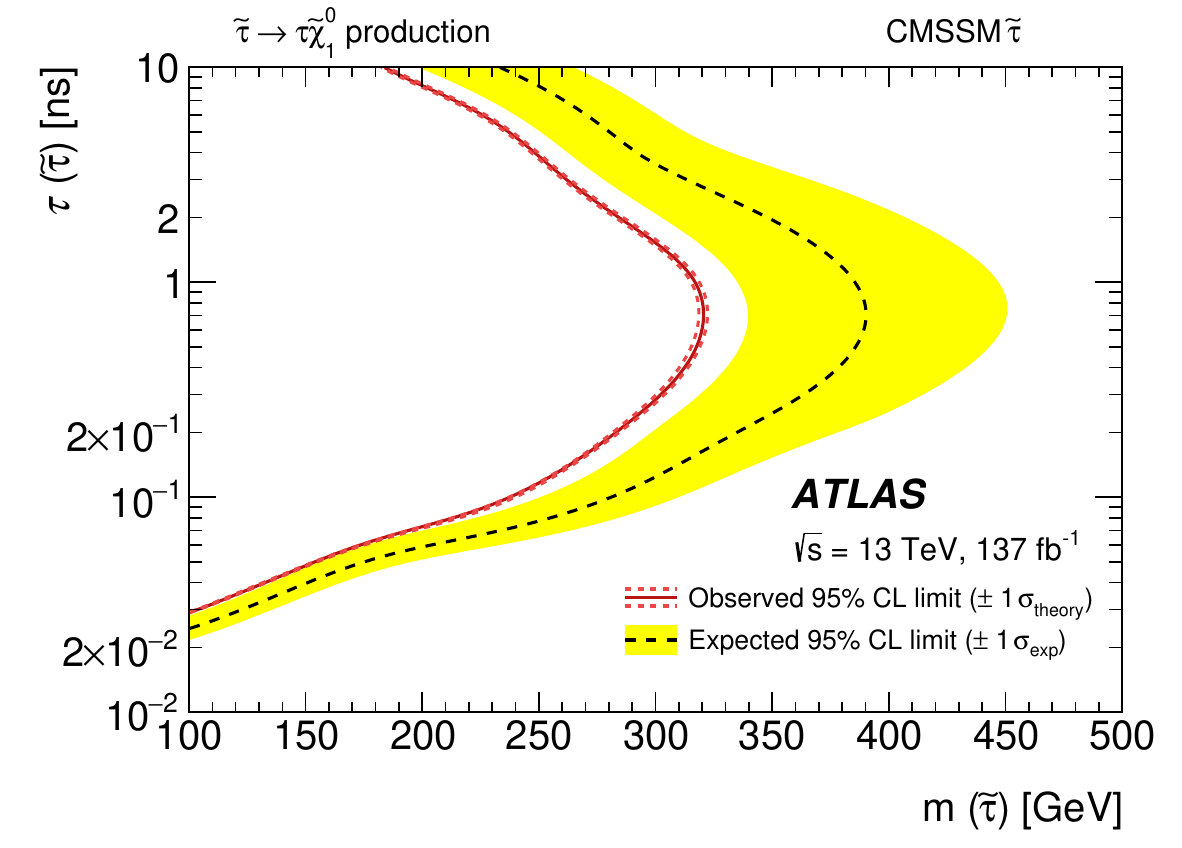}
\label{fig:results:cmssm_stau}
}
\subfigure[]{
\includegraphics[width=0.45\textwidth]{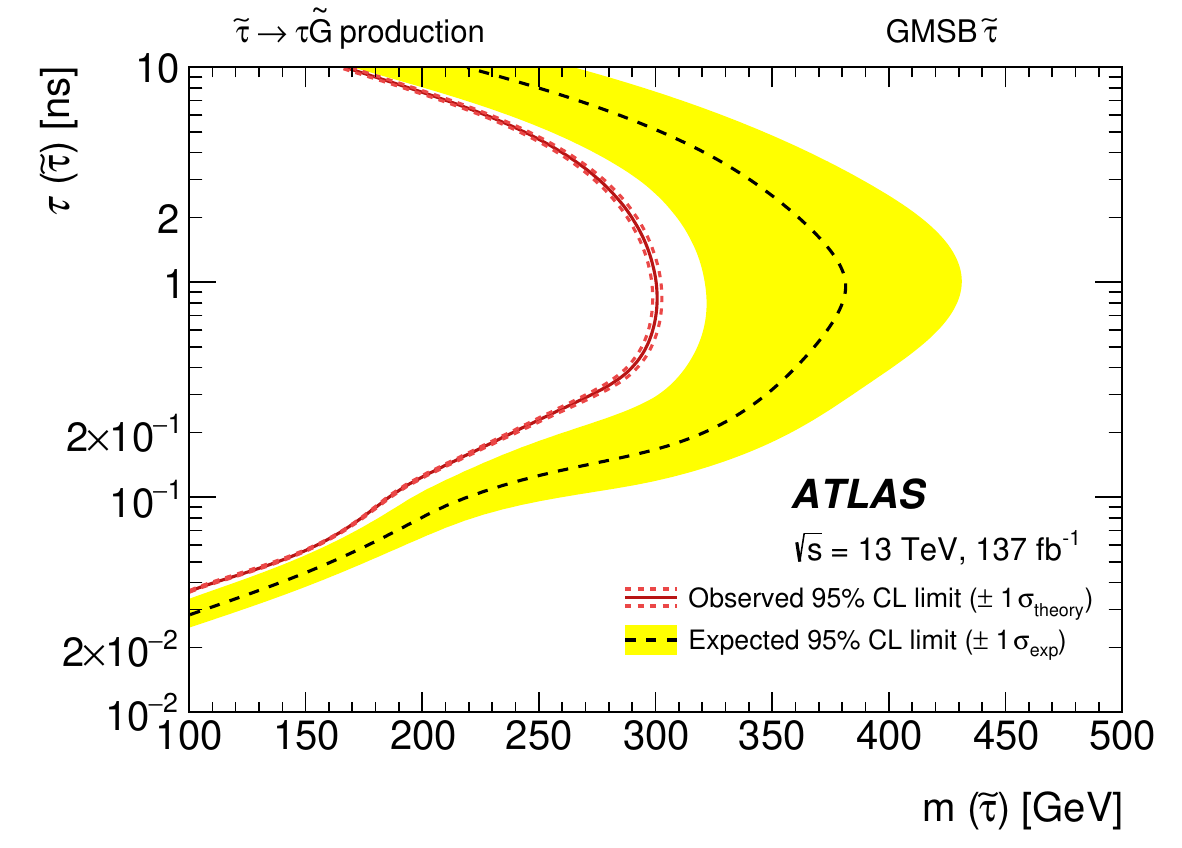}
\label{fig:results:gmsb_stau}
}
\end{center}
\vspace{-0.75cm}
\caption{The expected (dashed line) and observed (solid line) 95\% CL exclusion contours for $\tau$-slepton production in \protect\subref{fig:results:cmssm_stau} CMSSM-inspired scenarios and \protect\subref{fig:results:gmsb_stau} GMSB-inspired scenario. The dotted lines indicate the $\pm \sigma_{\mathrm{theory}}$ from signal cross-section uncertainties, while the solid band indicates the $\pm \sigma_{\mathrm{exp}}$ from experimental systematic uncertainties and statistical uncertainties on the data yields.\label{fig:results:StauLimits}}
\end{figure}

Due to changes in the analysis strategy since the previous analysis in Ref.~\cite{SUSY-2018-19}, particularly in the reconstruction of the tracklets, a direct comparison of the SRs is non-trivial, as the previous analyses used five-layer tracking and did not use the additional SCT data quality veto. Despite this, it is found that one of the events selected by \SRAh was also one of the three events selected by the previous analysis. The two new events in \SRAh are selected because of improvements in track reconstruction.
The two events selected by \SRBo are likely to stem from a hadronic interaction in the ID material as a secondary vertex is reconstructed in a region with known material. The resulting kink in the particle path results in a reconstructed tracklet rather than a standard track. A \enquote{material veto} was considered in the analysis design to avoid such events, but was not adopted as it had a highly detrimental impact on the signal acceptance.

% End of text imported from the .//sections/results.tex input file

%
%
\FloatBarrier

\section{Conclusion}
\label{sec:conclusion}
%

% The next lines are included from the .//sections/conclusion.tex input file
A search for new physics in final states containing a disappearing track and \met, indicative of a heavy charged particle decaying after traversing only the innermost layers of the ATLAS detector, has been performed using 137\,fb$^{-1}$ of data from $\sqrt{s} = 13\,\TeV$ $pp$ collisions at the LHC. Two BSM scenarios are considered, the first being electroweakino pair production with at least one chargino, where $\chinoonepm \rightarrow  \pi^{\pm}\ninoone$. The second set of scenarios considered are those of $\tau$-slepton pair production: a CMSSM-inspired scenario where the $\tau$-slepton decays as $\tilde{\tau} \rightarrow \tau \ninoone$, and a GMSB-inspired scenario where it decays as $\tilde{\tau} \rightarrow \tau \tilde{G}$. The analysis uses four signal regions for a final state with zero leptons, one jet, high \met, and a tracklet. Improvements in the analysis techniques, including the reconstruction of tracklets from hits in only three pixel layers and a dedicated BDT to reconstruct low-energy pions, increased the sensitivity compared to the previous Run-2 analysis. Data-driven methods are used to estimate the backgrounds, which arise when a stable particle is incorrectly reconstructed as a tracklet or when random combinations of inner-detector hits are incorrectly identified as a tracklet.

The data observations agree with the SM expectation to within $2\sigma$. The results are interpreted in simplified SUSY models. For higgsino production, chargino masses of up to $225\,\GeV$ ($250\,\GeV$) are observed (expected) to be excluded at 95\% CL in the lifetime regime $\tau < 0.03$\,ns where the electroweakino mass splitting is motivated by loop corrections due to SM bosons. These are currently the strongest exclusion limits in this region of phase space. For longer lifetimes up to $1$\,ns, exclusion limits of up to $720\,\GeV$ ($840\,\GeV$) are set. For wino production, charginos with masses of up to $880\,\GeV$ ($1020\,\GeV$) are excluded for lifetimes of around $1\,$ns. In CMSSM- and GMSB-inspired $\tau$-slepton production scenarios, the observed (expected) $\tau$-slepton mass exclusion limits reach $320\,\GeV$ ($390\,\GeV$) and $300\,\GeV$ ($380\,\GeV$) respectively, in the case of $\tau$-slepton lifetimes of around $1\,$ns.

% End of text imported from the .//sections/conclusion.tex input file

%
%
%
%
%
\section*{Acknowledgements}
%

%

% The next lines are included from the .//acknowledgements/Acknowledgements.tex input file
%
%

%
%

We thank CERN for the very successful operation of the LHC and its injectors, as well as the support staff at
CERN and at our institutions worldwide without whom ATLAS could not be operated efficiently.

The crucial computing support from all WLCG partners is acknowledged gratefully, in particular from CERN, the ATLAS Tier-1 facilities at TRIUMF/SFU (Canada), NDGF (Denmark, Norway, Sweden), CC-IN2P3 (France), KIT/GridKA (Germany), INFN-CNAF (Italy), NL-T1 (Netherlands), PIC (Spain), RAL (UK) and BNL (USA), the Tier-2 facilities worldwide and large non-WLCG resource providers. Major contributors of computing resources are listed in Ref.~\cite{ATL-SOFT-PUB-2026-001}.

We gratefully acknowledge the support of ANPCyT, Argentina; YerPhI, Armenia; ARC, Australia; BMWFW and FWF, Austria; ANAS, Azerbaijan; CNPq and FAPESP, Brazil; NSERC, NRC and CFI, Canada; CERN; ANID, Chile; CAS, MOST and NSFC, China; Minciencias, Colombia; MEYS CR, Czech Republic; DNRF and DNSRC, Denmark; IN2P3-CNRS and CEA-DRF/IRFU, France; SRNSFG, Georgia; BMFTR, HGF and MPG, Germany; GSRI, Greece; RGC and Hong Kong SAR, China; ICHEP and Academy of Sciences and Humanities, Israel; INFN, Italy; MEXT and JSPS, Japan; CNRST, Morocco; NWO, Netherlands; RCN, Norway; MNiSW, Poland; FCT, Portugal; MNE/IFA, Romania; MSTDI, Serbia; MSSR, Slovakia; ARIS and MVZI, Slovenia; DSI/NRF, South Africa; MICIU/AEI, Spain; SRC and Wallenberg Foundation, Sweden; SERI, SNSF and Cantons of Bern and Geneva, Switzerland; NSTC, Taipei; TENMAK, T\"urkiye; STFC/UKRI, United Kingdom; DOE and NSF, United States of America.

Individual groups and members have received support from BCKDF, CANARIE, CRC and DRAC, Canada; CERN-CZ, FORTE and PRIMUS, Czech Republic; COST, ERC, ERDF, Horizon 2020 and Marie Sk{\l}odowska-Curie Actions, European Union; Investissements d'Avenir Labex, Investissements d'Avenir Idex and ANR, France; DFG and AvH Foundation, Germany; Herakleitos, Thales and Aristeia programmes co-financed by EU-ESF and the Greek NSRF, Greece; BSF-NSF and MINERVA, Israel; NCN and NAWA, Poland; La Caixa Banking Foundation, CERCA and AGAUR programs from Generalitat de Catalunya and PROMETEO and GenT Programmes Generalitat Valenciana, Spain; G\"{o}ran Gustafssons Stiftelse, Sweden; The Royal Society and Leverhulme Trust, United Kingdom; Eric and Wendy Schmidt Fund for Strategic Innovation, United States of America.

In addition, individual members wish to acknowledge support from Chile: Agencia Nacional de Investigaci\'on y Desarrollo (ANID FONDECYT reg. 1230987, FONDECYT 1230812, FONDECYT 1240864, Fondecyt 3240661, Fondecyt Regular 1240721); China: Chinese Ministry of Science and Technology (MOST-2023YFA1605700, MOST-2023YFA1609300), National Natural Science Foundation of China (NSFC 12275265, NSFC-W2543005); Czech Republic: Czech Science Foundation (GACR - 24-11373S), Ministry of Education Youth and Sports (ERC-CZ-LL2327, FORTE CZ.02.01.01/00/22\_008/0004632), PRIMUS Research Programme (PRIMUS/21/SCI/017); EU: H2020 European Research Council (ERC - 101002463); European Union: European Research Council (BARD No. 101116429, ERC - 948254, ERC 101089007), European Regional Development Fund (HE COFUND GA No.101081355, ERDF), Marie Sklodowska-Curie Actions (GAP-101168829); France: Agence Nationale de la Recherche (ANR-21-CE31-0013, ANR-22-EDIR-0002, ANR-24-CE31-0504-01); Germany: Deutsche Forschungsgemeinschaft (DFG - 469666862); China: Research Grants Council (GRF); Italy: Ministero dell'Università e della Ricerca (NextGenEU 153D23001490006 M4C2.1.1, NextGenEU I53D23000820006 M4C2.1.1, NextGenEU I53D23001490006 M4C2.1.1, SOE2024\_0000023); Japan: Japan Society for the Promotion of Science (JSPS KAKENHI  JP25H0063, JSPS KAKENHI JP22H01227, JSPS KAKENHI JP22H04944, JSPS KAKENHI JP22KK0227, JSPS KAKENHI JP24K23939, JSPS KAKENHI JP24KK0251, JSPS KAKENHI JP25H00650, JSPS KAKENHI JP25H01291, JSPS KAKENHI JP25K01011, JSPS KAKENHI JP25K01023); Norway: Research Council of Norway (RCN-314472); Poland: Polish National Science Centre (NCN 2021/42/E/ST2/00350, NCN OPUS 2023/51/B/ST2/02507, NCN OPUS nr 2022/47/B/ST2/03059, NCN UMO-2019/34/E/ST2/00393, UMO-2022/47/O/ST2/00148, UMO-2023/49/B/ST2/04085, UMO-2023/51/B/ST2/00920, UMO-2024/53/N/ST2/00869); Spain: Agència de Gestió d'Ajuts Universitaris i de Recerca. (AGAUR - 2023 BP 00141), Ministry of Science and Innovation (RYC2019-028510-I, RYC2020-030254-I, RYC2021-031273-I, RYC2022-038164-I), Ministerio de Ciencia, Innovación y Universidades/Agencia Estatal de Investigaci\'on (EU NextGenerationEU (PRTR-C17.I1), PID2022-142604OB-C22); Sweden: Carl Trygger Foundation (Carl Trygger Foundation CTS 22:2312), Swedish Research Council (Swedish Research Council 2023-04654, VR 2021-03651, VR 2022-03845, VR 2022-04683, VR 2023-03403, VR 2024-05451, VR 2025-05940), Knut and Alice Wallenberg Foundation (KAW 2023.0366); Switzerland: Swiss National Science Foundation (SNSF - PCEFP2\_194658); United Kingdom: The Binks Trust, Royal Society (NIF-R1-231091); United States of America: U.S. Department of Energy (ECA DE-AC02-76SF00515), John Templeton Foundation (John Templeton Foundation 63206), Neubauer Family Foundation.

%
%

% End of text imported from the .//acknowledgements/Acknowledgements.tex input file

%
\clearpage
\printbibliography
\clearpage
\input{atlas_authlist}
%

%
%

%

%
%
%
%

\end{document}

%% file: atlas_authlist.tex
% ATLAS Collaboration author list
% Reference date of HMBS-2024-66 is 2026-01-30
% Author list last updated on date 08-MAY-26
% Data extracted on 08-May-2026 for paper reference HMBS-2024-66
% at 10:37am
 
\begin{flushleft}
\hypersetup{urlcolor=black}
{\Large The ATLAS Collaboration}

\bigskip

\AtlasOrcid[0000-0002-6665-4934]{G.~Aad}$^\textrm{\scriptsize 102}$,
\AtlasOrcid[0000-0001-7616-1554]{E.~Aakvaag}$^\textrm{\scriptsize 17}$,
\AtlasOrcid[0000-0002-5888-2734]{B.~Abbott}$^\textrm{\scriptsize 121}$,
\AtlasOrcid[0000-0002-0287-5869]{S.~Abdelhameed}$^\textrm{\scriptsize 83b}$,
\AtlasOrcid[0000-0002-1002-1652]{K.~Abeling}$^\textrm{\scriptsize 54}$,
\AtlasOrcid[0000-0001-5763-2760]{N.J.~Abicht}$^\textrm{\scriptsize 48}$,
\AtlasOrcid[0000-0002-8496-9294]{S.H.~Abidi}$^\textrm{\scriptsize 30}$,
\AtlasOrcid[0009-0003-6578-220X]{M.~Aboelela}$^\textrm{\scriptsize 44}$,
\AtlasOrcid[0000-0002-9987-2292]{A.~Aboulhorma}$^\textrm{\scriptsize 36e}$,
\AtlasOrcid[0000-0001-5329-6640]{H.~Abramowicz}$^\textrm{\scriptsize 154}$
% ...
\bigskip
\\

% ...
$^{17}$Department for Physics and Technology, University of Bergen, Bergen; Norway.\\
% ...
$^{30}$Physics Department, Brookhaven National Laboratory, Upton NY; United States of America.\\
% ...
$^{36}$$^{(a)}$Facult\'e des Sciences Ain Chock, Universit\'e Hassan II de Casablanca;$^{(b)}$Facult\'{e} des Sciences, Universit\'{e} Ibn-Tofail, K\'{e}nitra;$^{(c)}$Facult\'e des Sciences Semlalia, Universit\'e Cadi Ayyad, LPHEA-Marrakech;$^{(d)}$LPMR, Facult\'e des Sciences, Universit\'e Mohamed Premier, Oujda;$^{(e)}$Facult\'e des sciences, Universit\'e Mohammed V, Rabat;$^{(f)}$Institute of Applied Physics, Mohammed VI Polytechnic University, Ben Guerir; Morocco.\\
% ...
$^{44}$Physics Department, Southern Methodist University, Dallas TX; United States of America.\\
% ...
$^{48}$Fakult\"{a}t Physik, Technische Universit{\"a}t Dortmund, Dortmund; Germany.\\
% ...
$^{54}$II. Physikalisches Institut, Georg-August-Universit\"{a}t G\"ottingen, G\"ottingen; Germany.\\
% ...
$^{83}$$^{(a)}$Khalifa University of Science and Technology, Abu Dhabi;$^{(b)}$New York University Abu Dhabi, Abu Dhabi;$^{(c)}$United Arab Emirates University, Al Ain;$^{(d)}$University of Sharjah, Sharjah; United Arab Emirates.\\
% ...
$^{102}$CPPM, Aix-Marseille Universit\'e, CNRS/IN2P3, Marseille; France.\\
% ...
$^{121}$Homer L. Dodge Department of Physics and Astronomy, University of Oklahoma, Norman OK; United States of America.\\
% ...
$^{154}$Raymond and Beverly Sackler School of Physics and Astronomy, Tel Aviv University, Tel Aviv; Israel.\\
% ...
$^{a}$ Also at Affiliated with an institute formerly covered by a cooperation agreement with CERN.\\
$^{b}$ Also at An-Najah National University, Nablus; Palestine.\\
$^{c}$ Also at Borough of Manhattan Community College, City University of New York, New York NY; United States of America.\\
$^{d}$ Also at Center for Interdisciplinary Research and Innovation (CIRI-AUTH), Thessaloniki; Greece.\\
$^{e}$ Also at Centre of Physics of the Universities of Minho and Porto (CF-UM-UP); Portugal.\\
$^{f}$ Also at CERN, Geneva; Switzerland.\\
% ...

\end{flushleft}

% Created with Glance <Atlas.Glance@cern.ch>